\documentclass{aastex}
\usepackage{emulateapj5}

\shortauthors{Simard et al.}
\shorttitle{HST Structural Parameters of GSS Galaxies}

\def\deg{\mbox{$^\circ$}}

\def\littleprime{\ifmmode{\scriptscriptstyle \prime }   
\else{\hbox{$\scriptscriptstyle \prime$ }}\fi}

\def\arcsec{\raise .9ex \hbox{\littleprime\hskip-3pt\littleprime}}
\def\arcmin{\raise .9ex \hbox{\littleprime}}
\def\arcsecpoint{\hbox to 1pt{}\rlap{\arcsec}.\hbox to 2pt{}}
\def\arcminpoint{\hbox to 1pt{}\rlap{\arcmin}.\hbox to 2pt{}}

\def\eg {\mbox{e. g.}}
\def\ie {\mbox{i. e.}}

\def\spose#1{\hbox to 0pt{#1\hss}}
\def\lta{\mathrel{\spose{\lower 3pt\hbox{$\mathchar"218$}}
     \raise 2.0pt\hbox{$\mathchar"13C$}}}
\def\gta{\mathrel{\spose{\lower 3pt\hbox{$\mathchar"218$}}
     \raise 2.0pt\hbox{$\mathchar"13E$}}}

\received{2002 January 17}
\accepted{2002 April 16}
\slugcomment{Accepted for publication in ApJ Supplements, 16 April 2002}
\begin{document}

\title{The DEEP Groth Strip Survey II. Hubble Space Telescope
Structural Parameters of Galaxies in the Groth Strip.\altaffilmark{1,2}}

\author{Luc Simard\altaffilmark{3,4}, Christopher N. A. 
Willmer\altaffilmark{5}, Nicole P. Vogt\altaffilmark{6}, Vicki L. 
Sarajedini\altaffilmark{7}, Andrew C. Phillips, Benjamin J. Weiner, David C. Koo, 
Myungshin Im\altaffilmark{8}, Garth D. Illingworth, S. M. Faber} 
\affil{UCO/Lick Observatory, University of California, Santa Cruz, CA 
95064, USA}

\altaffiltext{1}{Based on observations made with the NASA/ESA {\it 
Hubble Space Telescope} which is operated by AURA, Inc., under 
contract with NASA.}

\altaffiltext{2}{Based on data obtained at 
the W.M. Keck Observatory, which is operated as a scientific 
partnership among the California Institute of Technology, the 
University of California and the National Aeronautics and Space 
Administration.  The Observatory was made possible by the generous 
financial support of the W.M. Keck Foundation.}

\altaffiltext{3}{Guest User, Canadian Astronomy Data Center, which is 
operated by the National Research Council of Canada, Herzberg 
Institute of Astrophysics, Dominion Astrophysical Observatory} 

\altaffiltext{4}{Current Address: Herzberg Institute of Astrophysics, National 
Research Council of Canada, 5071 West Saanich Road, Victoria, BC V9E 2E7, 
Canada}

\altaffiltext{5}{On leave from Observatorio Nacional, Rua General Jose 
Cristino 77, 20921-030 Sao Cristovao, RJ, Brazil}

\altaffiltext{6}{Current Address: Department of Astronomy, New Mexico State 
University, Las Cruces, NM 88003-8001, USA}

\altaffiltext{7}{Current Address: Department of Astronomy, University of Florida, 
Gainesville, FL 32611-2055, USA}

\altaffiltext{8}{Current Address: SIRTF Science Center/IPAC, Mail Stop 220-6, 
California Institute of Technology, Pasadena, CA 91125, USA}

\begin{abstract}
The quantitative morphological classification of distant galaxies is essential to 
the understanding of the evolution of galaxies over the history of the Universe.  
This paper presents {\it Hubble Space Telescope} WFPC2 $F606W$ and 
$F814W$ photometric structural parameters for 7450 galaxies in the ``Groth Strip.'' 
These parameters are based on a two-dimensional bulge+disk surface 
brightness model and were obtained using an automated reduction and analysis pipeline described 
in detail here.  A first set of fits was performed separately in each  bandpass, 
and a second set of fits was performed simultaneously on both bandpasses. 
The information produced by these two types of fits can be used to explore different 
science goals.  Systematic and random fitting errors in all structural parameters as 
well as bulge and disk colors are carefully characterized through extensive 
sets of simulations.  The results of these simulations are given in catalogs similar 
to the real science catalogs so that both real and simulated measurements  can be 
sampled according to the same selection criteria to show biases and errors in the 
science data subset of interest.  The effects of asymmetric structures on the recovered 
bulge+disk fitting parameters are also explored through simulations.  The 
full multidimensional photometric survey selection function of the Groth Strip is 
also computed.  This selection function, coupled to bias maps from simulations, 
provides a complete and objective reproduction of the observational limits, and these limits can 
be applied to theoretical predictions from galaxy evolution models for direct 
comparisons with the data.  
\end{abstract}


\section{Introduction} \label{intro}

\noindent The visual classification of galaxies has a venerable 
tradition in optical astronomy starting with the introduction of 
Hubble's famous ``Tuning Fork'' diagram \citep{hubble36}.  Despite the 
fact that others have extended Hubble's original classification mainly 
to deal with the diversity of structures in later-type galaxies \citep{devauc59,vdb60a, vdb60b, vdb76,morgan70}, the 
nomenclature of Hubble still very much pervades the language of 
today's galaxy morphology literature.  This longevity is a tribute to 
Hubble's seminal work.  However, visual classification has serious 
weaknesses.  First and foremost, it is a subjective process.  Although 
visual classification experts can agree to within two revised Hubble 
types \citep{naim95}, it is very difficult for non-experts to produce 
reliable visual classifications without years of hard-won experience.  
Second, it is also unclear how useful visual classification is with 
respect to high-redshift galaxies.  Limited spatial resolution means 
that larger and larger internal galaxy structures such as spiral arms 
and tidal tails get progressively smoothed out with increasing 
redshift, and this smoothing can introduce significant classification 
biases.

No visual or quantitative classification system is perfect.  However, 
provided a given quantitative system is clearly defined by its 
proponents, it is readily reproducible in its successes {\it and} 
failures by others.  This is the fundamental advantage of the 
quantitative approach.  Moreover, systematic and random errors of 
quantitative classifiers can be carefully characterized through 
extensive sets of galaxy image simulations covering a wide range of 
structural parameters.  This is another important advantage.  It 
should be emphasized that quantitative classification is meaningless 
without three important elements: the measurements themselves, the 
simulations and the galaxy selection function.  This selection 
function is critical to relate observed structural parameter 
distributions to predictions from theoretical models.

A number of quantitative classifiers have been developed/extended over 
the years to probe the structure of high-redshift galaxies.  These 
classifiers can be either parametric (model-based) or non-parametric.  
Non-parametric classifiers include the $C-A$ system 
\citep{watanabe85,abraham94, abraham96,wu99,conselice00}, artificial neural 
nets trained from visual classification sets \citep{odewahn95,odewahn96}, and self-organizing maps \citep{naim97}.  
Parametric classifiers include radial multi-gaussian deconvolution 
\citep{bendinelli91,fasano98} and bulge+disk decomposition 
\citep{schade95, schade96,ratnatunga99}.  The latter is 
popular for three reasons: (1) It is rooted in the very first studies 
of the functional form of galaxy radial surface brightness profiles 
\citep{devauc48, devauc59}; (2) It provides a ``comfortable'' mental 
picture of the overall structure of a distant galaxy, \ie, it is 
conceptually simple to relate a quantitative measurement of galaxy 
type such as bulge-to-total light ratio to the familiar Hubble types; 
and (3) Photometric entities such as bulges and disks have distinct 
dynamical counterparts although this correspondence does not always 
hold (see Section \ref{surfmodel} for more details).

This work uses GIM2D (Galaxy IMage 2D), a 2D decomposition fitting 
program \citep[][and this paper]{simard98}, to measure the structural 
parameters of galaxies in the ``Groth Strip'' \citep{groth94,rhodes00}.  GIM2D 
is an IRAF\footnote{IRAF is distributed by the National Optical 
Astronomy Observatories, which are operated by the Association of 
Universities for Research in Astronomy, Inc., under cooperative 
agreement with the National Science Foundation.}/SPP package written 
to perform detailed bulge+disk surface brightness profile 
decompositions of low signal-to-noise (S/N) images of distant galaxies 
in a fully automated way.  GIM2D takes an HST or ground-based science 
image and its source catalog, performs 2D profile fits on each source 
and produces model-subtracted images as well as a full catalog of 
structural parameters.  The bulge+disk model adopted here is not 
fundamentally new, but GIM2D offers an independent check of other 
galaxy classification works by including a set of extended features 
(S\'ersic bulge profile, a comprehensive but by all means not 
exhaustive set of image asymmetry indices, three different fitting 
methods) and a different fitting algorithm.  GIM2D has already been 
used in a variety of HST and ground-based distant galaxy studies: the 
optical structure of intermediate redshift compact narrow-emission line 
galaxies \citep{guzman98}, the quantitative morphology of Hubble Deep Field North 
galaxies \citep{marleau98}, the NICMOS structure of a spiral galaxy 
lens at $z$ = 0.4 \citep{maller00}, the luminosity-size relation of 
field disk galaxies from $z = 0.1$ to $z = $ 1.1 \citep{simard99}, 
the number density and luminosity function of E/SO galaxies to $z 
\lesssim 1$ \citep{im01}, the Fundamental Plane of field absorption-line 
galaxies out to $z \sim 1$ \citep{gebhardt02}, tests of hierarchical 
galaxy evolution models \citep{simard02}, the colors of luminous 
bulges at $z \sim 1$ \citep{koo02},  the galaxy populations of poor, X-ray selected 
groups of galaxies \citep{tran01} and of high and low X-ray luminosity 
galaxy clusters \citep{balogh02}, and the size evolution of high-redshift brightest 
cluster galaxies \citep{nelson02}.

This paper describes in detail the structural parameter analysis of 
galaxies in the Groth Strip from calibrated HST archival images to 
final, parameter-rich structural catalogs for the entire Strip.  It is 
a companion paper to \citet{vogt02} and \citet{phillips02} 
in which the spectroscopic Keck/Low Resolution Imaging Spectrograph 
survey of the Groth Strip undertaken by the Deep Extragalactic 
Evolutionary Probe (DEEP) team is described.  Cosmological parameters 
adopted throughout this paper are $H_{0} = 70$ km s$^{-1}$ Mpc$^{-1}$, 
and ($\Omega_{m}$, $\Omega_{\Lambda}$, $\Omega_{k}$) = (0.3, 0.7, 
0.0).

\section{Overview} \label{overview}

Quantitative galaxy morphology classifiers must include three vital 
elements in order to be scientifically useful: the structural 
measurements themselves, extensive simulations, and a detailed survey 
selection function.

Every structural catalog should have a companion catalog of 
simulations from which the simulated counterparts of real data plots 
can be extracted.  The ability to clearly visualize the systematic and 
random errors of a set of observed structural parameters is very 
helpful to quickly assess what science goals can be effectively 
pursued with the measurements.  Morever, the systematic and random 
errors as characterized through these simulations can be applied to 
the theoretical predictions from galaxy evolution models to 
``scatter'' the models in the same way as the observational errors 
would scatter signals from the real Universe.

The survey selection function is constructed by inserting galaxy 
images with a wide range of input structural parameters into real 
images and calculating the success rate of the detection algorithm 
as a function of those parameters. This selection function serves two 
purposes: (1) It can be used to insure that studies of distant 
galaxies spanning a large range of redshifts treat galaxy samples at 
different redshifts in the same way (\citealp[e.g.,][]{simard99}). (2) 
The selection function can be used to ``observe'' theoretical models 
for a direct comparison with the data (\citealp[e.g.,][]{simard02}). 
Together with the simulations, the selection function provides the best 
possible way to reproduce the biases of the observational strategy.

Current and planned future on-line archival data systems contain (or 
will contain) prodigious amounts of data.  Mining these systems in as 
automated a mode as possible holds the promise of fantastic 
scientific returns on problems requiring very large, statistically 
well-defined samples such as the evolution of galaxies.  The automated 
GIM2D pipeline described in the next sections was designed to produce 
the above three elements with data mining of large datasets in mind.

The GIM2D pipeline includes the following steps: (1) Pre-processing, cosmic 
ray (CR) rejection and data quality mask creation (Section~\ref{preproc}), 
(2) Source detection, deblending and extraction (Section~\ref{sext}), 
(3) Bulge+disk decompositions of galaxy images (Section~\ref{surfits}) 
with a Point-Spread-Function (PSF) for each object 
(Section~\ref{psf}), (4) Creation of residual images and computation 
of residual image indices (Section~\ref{resindices}), (5) Construction 
of measured structural parameter catalogs (Section~\ref{catalogs}), 
(6) Construction of associated catalogs of simulations for mapping systematic 
biases and random errors (Section~\ref{simulations}), and (7) 
Construction of the survey selection function (Section~\ref{selecf}).

\section{Pre-processing of HST images}\label{preproc}

The GIM2D pipeline starts with the retrieval and pre-processing of HST 
archival images.  Archival images are calibrated ``on-the-fly'' at 
retrieval time.  The retrieved images are then sent through a set of 
IRAF scripts that removes cosmic rays and combine all the images taken 
in the same bandpass and at the same position on the sky together.  
The pipeline does not mosaic the three WFC chips together.  Mosaicing 
(\eg, IRAF/STSDAS/WMOSAIC) destroys the uncorrelated noise 
characteristics of the original images by applying flux interpolations 
and geometric corrections.  The noise characteristics of the science 
images must be preserved as much as possible through the whole 
pipeline so that it is possible to provide a realistic noise estimate 
to the likelihood function during the galaxy image fitting process.  
Each WFPC2 pointing is therefore split into three different images, 
one for each Wide-Field chip, and each chip is then processed 
independently through the rest of the pipeline.  Each WFC chip covers a 
80\arcsec $\times$ 80\arcsec field of view with a pixel scale of 0\arcsecpoint1/pixel.
The Planetary Camera images were not analyzed since all the DEEP/GSS 
Keck spectroscopic targets were selected solely from the WFC images 
for the sake of sample homogeneity.

\subsection{HST/WFPC2 Image Datasets}\label{hstimages}

The HST images are from two surveys, collectively dubbed the ``Groth 
Survey Strip'' (GSS), taken under HST programs GTO 5090 (PI: Groth) 
and GTO 5109 (PI: Westphal).  The GSS consists of 28 overlapping 
subfields taken with the HST Wide Field and Planetary Camera 2 (WFPC2) 
and forms a ``chevron strip'' oriented NE to SW at roughly 1417+52 at 
Galactic latitude $b \sim 60\deg$.  Each of 27 subfields has exposures 
of 2800 seconds (4$\times$ 700 seconds) in the broad $V$ filter ($F606W$) and 4400 
seconds (4 $\times$ 1100 seconds) in the broad $I$ filter ($F814W$).  The 28th 
field is the Westphal Deep Survey Field 2 (J2000 1417.5+52.5), with 
total exposures of 24,400 seconds in $V$ and 25,200 seconds in $I$.  The images 
were recalibrated ``on-the-fly'' through the Canadian Astronomy Data 
Center (CADC) standard pipeline \citep{pirenne98}.  The datasets 
analyzed in this paper are listed in Table~\ref{hst-images-lst}.  
Field 7 is the Westphal Deep Survey Field.

\subsection{WFPC2 Data Quality Frames}\label{dqfframes}
Archival data produced through on-the-fly recalibration come in two 
parts: the science frames and the data quality frames (DQF).  Each 
science frame comes with a data quality frame.  This frame flags 
pixels in the science frame that have been corrupted for one reason or 
another (see Section 26.2.2 of HST Data Handbook Version 3.0 for DQF 
flag values).  A DQF flag value of zero is used by the HST pipeline to 
flag good pixels.  The GIM2D pipeline goes through the data quality 
frames of all science images in a given image stack\footnote{An image 
stack is defined here as a set of consecutive exposures taken at the 
same location on the sky and through the same filter.} and identifies 
all the pixels with no good pixel values in the stack.  The locations 
of these pixels are recorded and applied to the SExtractor 
segmentation image (Section~\ref{deblend}) to make sure bad pixels are 
not included in the surface brightness fits.

\subsection{Cosmic Ray Rejection and Image Combination}\label{coscomb}

The IRAF/STSDAS task CRREJ was used to reject cosmic rays from GSS 
WFPC2 image stacks.  It combines a stack of consecutive exposures 
while eliminating cosmic rays and scaling the remaining pixels to the 
total exposure time of the stack.  Cosmic rays are rejected through a 
series of sigma-clipping iterations.  The number and sigma thresholds 
of these iterations are specified using the parameter SIGMAS. SIGMAS 
must be chosen carefully.  If SIGMAS is too high, too many cosmic rays 
will be missed.  On the other hand, if SIGMAS is too low, the CR 
rejection algorithm will start eating away at the noise in the 
background pixels.  Unfortunately, there is no prescribed way of 
determining SIGMAS. The approach adopted here involved first the 
creation of a combined image with such high SIGMAS values that no 
pixels were rejected by CRREJ. The resulting ``HIGH'' image showed all 
the cosmic rays that hit all the science images in the image stack.  
However, there were also plenty of untouched background pixels that 
showed what the background in the combined image should look like.  
CRREJ-combined images were then created with progressively lower 
values of SIGMAS, and blinked against the ``HIGH'' image.  SIGMAS 
values were lowered until the noise in the untouched background areas 
of these images clearly showed they were being modified by the 
sigma-clipping with respect to the same areas in the ``HIGH'' image.  
The best value of SIGMAS was found to be ``6,4'' \ie, CRREJ was 
instructed to perform two sigma-clipping iterations, the first one at 
the $6\sigma$-level and the second one at the $4\sigma$-level.  This 
value was then automatically applied in the combinations of all the 
GSS stacks.  Even with the best SIGMAS cuts, a number of low-energy 
cosmic rays will be left in the final combined image.  If these 
low-energy cosmic rays significantly changed the background noise 
properties of the final combined images, the background pixel 
histograms of these images should show deviations from a Gaussian 
distribution in the form of high flux tails.  No such deviations were 
ever observed in the histograms inspected at different background 
locations in the final images.

\subsection{WFPC2/GSS Photometric Zeropoints}\label{photzero}

All total $F814W$ fluxes $F_{814}$ (galaxy, bulge or disk) 
will be converted in this paper to $F814W$ magnitudes on the Vega system 
(denoted $I_{814}$ or simply $I$ hereafter) using the equation:

\begin{equation}
	I_{814} = -2.5\thinspace {\rm log}_{10}(F_{814}/t_{814}) + C_{814}, 
	\label{magdefI}
\end{equation}
\noindent where $C_{814}$ = 21.65 (May 1997 WFPC2 
SYNPHOT update).  

Similarly, all total $F606W$ fluxes $F_{606}$ (galaxy, bulge or disk) 
were converted to $F606W$ magnitudes on the Vega system (denoted 
$V_{606}$ or simply $V$ hereafter) using the equation:

\begin{equation}
	V_{606} = -2.5\thinspace {\rm log}_{10}(F_{606}/t_{606}) + C_{606}, 
	\label{magdefV}
\end{equation}
\noindent where $C_{606}$ = 22.91 (May 1997 WFPC2 
SYNPHOT update).  The total exposure time $t_{814}$ was 4400 seconds in the 
$F814W$ filter, and the total exposure time $t_{606}$ was 2800 seconds in the 
$F606W$ filter in all GSS fields except Field 7 for which $t_{814}$ = 25200 
seconds and $t_{606}$ = 24400 seconds.

\section{Source Detection, Deblending and Extraction}\label{sext}

To proceed with the fitting of galaxy images, GIM2D needs a catalog of 
sources for each image to be analyzed.  For each source the catalog 
must include a $x$-$y$ centroid position, an initial estimate of the local sky 
background level and the isophotal area of the object in pixels above 
the detection threshold.  GIM2D also needs a segmentation or mask 
image in which pixels belonging to the same object are all assigned 
the same flag value and sky background pixels are flagged by zeroes.  
The source catalogs and segmentation images for the Groth Strip were 
created using the SExtractor (``Source Extractor'') galaxy photometry 
package version 1.0a \citep{bertin96}.

\subsection{Detection Parameters}\label{sexdetect}

The SExtractor source detection was run on the CRREJ-combined 
$I_{814}$ GSS images.  The detection threshold was 1.5-$\sigma_{bkg}$, 
and the required minimum object area above that threshold was 10 
pixels.  The convolution kernel was a 3$\times$3 Gaussian kernel with 
a FWHM of 1.5 pixels.  These detection parameters do not have to be 
particularly fine-tuned to extract the faintest possible sources from 
the GSS images since the faintest magnitude at which reliable 
bulge+disk decompositions can be performed is well above the 
magnitude limits of the SExtractor source catalog.  No star/galaxy 
separation was attempted.  Every source was fitted with GIM2D. 
Unresolved sources such as stars could easily be identified as GIM2D 
output models with zero half-light radius.

\subsection{Deblending}\label{deblend}

As SExtractor performs source detection and photometry, it is able to 
deblend sources using flux multi-thresholding.  This deblending 
technique works well in the presence of saddle points in the light 
profiles between objects.  Each SExtractor pre-deblending ``object'' 
consists of all the pixels above the detection threshold that are 
spatially connected to one another.  This group of pixels may or may not 
include several real objects.  For each ``object,'' the 
multi-thresholding algorithm goes through its connected pixels and 
rethresholds them at $N$ levels ($N = 32$ in the current analysis) 
between the 1.5$\sigma_{bkg}$ isophote and the peak pixel value in the 
``object'' to build a two-dimensional flux tree of the ``object.''  
The algorithm then goes down the threshold levels, and it looks at 
each branch in the tree to see if the flux contained in that branch 
above the threshold level is a fraction $f_{sep}$ or greater of the 
total flux in the ``object.''  If a given branch meets this condition, 
it is then treated as a separate object, and the separation threshold 
for that object is set to the threshold at which the split occurred.  
The multi-thresholding algorithm assigns the pixels between two 
adjacent objects and below the separation threshold based on a 
probability calculated from bivariate Gaussian fits to the two 
objects.  No assumption is made regarding the shape of the objects in 
this statistical deblending technique.

The fraction $f_{sep}$ is set by the SExtractor input parameter 
DEBLEND$_-$MINCONT. A value of 0.00075 was used for the SExtractor GSS 
source catalogs.  This value is {\it subjective}, and it was found 
through visual inspection of several GSS fields to provide good object 
separation.  Even though the value of DEBLEND$_-$MINCONT was 
determined subjectively, it provides an unequivocal definition of an 
object in the GSS catalogs presented in this paper.  It was only 
determined once, and the same value of DEBLEND$_-$MINCONT was 
consistently used for all GSS fields as well as for all GSS 
simulations.

Figure~\ref{gss-science-image} shows a typical section of the Groth 
Strip taken through the $F814W$ filter (DEEP/GSS Field ID 8/WFC Chip 
3), and Figure~\ref{gss-segmen-image} shows the corresponding 
SExtractor segmentation image.

\subsection{Thumbnail Image Extraction}\label{thumbim}

GIM2D disk+bulge decompositions are performed on thumbnail (or 
``postage stamp'') images extracted around the objects detected by 
SExtractor rather than on the entire science image itself.  Thumbnail 
images are preferable for two reasons: (1) Thumbnail images reduce the 
memory and I/O footprints of the program so it can be used in background 
mode on many computers without significantly impacting their other users.  
(2) Many CPUs can work on the same science image at the same time 
as they work on different thumbnail images.  This is extremely useful 
since the same list of thumbnail images can be sent to all available CPUs 
on a network/cluster, and all CPUs will work down the same master list 
without interfering with one another.  There is no limit on the number 
of CPUs that can be simultaneously harnessed for a given master list.

GIM2D will extract two or three thumbnail images for each object in the 
SExtractor catalog.  The area of an object's thumbnail images is given 
by the isophotal area of the object.  Here, all thumbnails were chosen 
to have an area 20 times larger than the 1.5-$\sigma_{bkg}$ isophotal area.  
The first thumbnail is extracted from the science image itself, and 
the local background calculated by SExtractor is subtracted from it so 
that it should have a background mean level close to zero.  The second 
thumbnail is extracted from the SExtractor segmentation image.  This 
segmentation thumbnail is modified so that bad pixels identified by 
the DQF frames (see section~\ref{dqfframes}) will be excluded from the 
surface brightness fits.  Some HST image datasets, most notably NICMOS 
datasets, include a ``sigma'' image giving the expected background + 
Poisson noise at each pixel.  If such a sigma image is available, 
GIM2D will automatically extract the third thumbnail image from it.

\section{Surface Brightness Fits}\label{surfits}

\subsection{Point-Spread-Functions}\label{psf}
GIM2D accepts four kinds of Point-Spread-Functions (PSFs): a 2D 
gaussian PSF, a delta function PSF, a user-given PSF or a TinyTim PSF. 
GIM2D automatically normalizes the total flux in all input PSFs to 1.0 
to ensure that this step has been performed.  The 2D gaussian PSF is 
generated by GIM2D with the seeing FWHM specified in the GIM2D 
parameter file.  If the delta function PSF option is selected, no 
PSF-convolution is performed on the galaxy image models.  PSF effects 
should not be important for structures spanning a large number of 
resolution elements.  Fits without PSF convolution require 
considerably less computation time, a definite advantage for very 
large objects.  The third kind of PSF is usually an image given by the 
user.  For example, this image could be a bright star extracted 
directly from the image or could be created from a set of PSF stars 
using a stellar photometry program such as DAOPHOT \citep{stetson87}.  
This option is particularly useful for ground-based studies in which 
sufficiently sampled PSF stars are easily found all over the science frames.  
TinyTim PSFs are generated by the Space Telescope Science Institute 
package {\sl TINY TIM} \citep{krist93} version 4.4.  TinyTim PSFs 
are available for all major imaging instruments onboard HST including 
WPFC2 and NICMOS. WFPC2 TinyTim PSFs were used for all 
(small and large) objects in the GSS analysis here.

The shape of the HST/WFPC2 PSF varies significantly as a function of 
position.  Therefore, $F814W$ and $F606W$ PSFs were generated with 
TinyTim for each GSS object analyzed here.  All GSS PSFs were 
oversampled by a factor of 5 and were 2\arcsecpoint 4 on a side.  
No telescope jitter was added.  The pointing stability of HST in fine 
lock mode is typically better than 5 mas RMS, and any value under 7 
mas is not noticeable \citep{krist93}.  WFPC2 charge-coupled devices (as 
any other CCDs) suffer from charge diffusion.  Charge diffusion 
contributes a small amount of blurring to the images, and TinyTim does 
include a WFPC2 subpixel response function to take charge diffusion 
into account.  However, it is very important to mention that TinyTim 
does {\it not} automatically apply this subpixel response function to 
oversampled PSFs.  GIM2D automatically convolves the oversampled
PSF-convolved galaxy model with the subpixel response function kernel 
given in \citet{krist93} after the model has been rebinned to the WFC 
detector resolution.

\subsection{Galaxy Image Model}\label{surfmodel}

The bulge+disk model used in GIM2D and other works is obviously a simple 
approximation.  After all, many real galaxies will exhibit more than two structural 
components such as nuclear sources, bars, spiral arms and HII regions.  Even in 
the presence of only a bulge and a disk, the ellipticity and/or the position angles of 
these components might be functions of galactocentric distance.  However, each 
new structural component or new layer of complexity added to the model comes 
with additional parameters that stretch the amount of information that can be reasonably 
extracted from small, low signal-to-noise images of distant galaxies.  The simple 
bulge+disk model is a trade-off between a reasonable number of fitting parameters 
and a meaningful decomposition of galaxy images.  Despite its relative simplicity, 
careful analysis of the parameters of the bulge+disk model can yield useful information 
regarding those higher layers of complexity.  For example, a very large deviation 
between the position angles of the bulge and of the disk is usually a strong indication 
of the presence of a bar, so barred galaxies can be easily identified without individually 
looking at all the galaxies in the sample.  Histograms of $\Delta\phi \equiv |\phi_b - 
\phi_d|$ are indeed a powerful way of finding barred galaxies. The shape of such 
histograms typically shows a well defined peak at $\Delta\phi$ = 0 with a tail of outliers 
that turn out to be the barred galaxies. The bulge+disk model is also a good way to 
study the morphology of quasar host galaxies. In the presence of a strong, unresolved, 
central source, the radius of one of the model components will shrink to zero to 
accommodate the source, and the other component will shape itself after the host galaxy. So, 
if quasar nuclei are found predominantly in elliptical galaxies, the radius of the disk 
component would be the one to converge to zero. Differences in the $F814W$ and 
$F606W$ centroid can also provide information on the irregularity of galaxy shape.  
All these examples illustrate the fact that the apparent simplicity of the bulge+disk 
model should not mask the richness of information provided by the inter-comparison 
of its full set of parameters.

Nonetheless, the bulge+disk model used here (and in other works) also has important 
limitations: it has a unique center for the whole galaxy,  and the intrinsic (\ie, before PSF 
convolution) ellipticity and position angle of each component do not change 
as a function of radius. This is in contrast to techniques such as isophotal ellipse 
fitting in which it is customary to let ellipticity, position angle and centroid vary from 
one ellipse to the next. The limitations of the bulge+disk model can yield results that are 
unexpected (or unsuspected!) at first glance. Consider a purely elliptical galaxy with no 
disk component whatsoever but with varying ellipticity and position angle with radius. 
A bulge+disk fit  to such an object may converge to a model with a significant disk 
component ($B/T < 1$). The additional degrees of freedom provided by the disk 
component are used by the model to compensate for the varying ellipticity and 
position angle. The position angle difference between the bulge and disk components 
will thus reflect the position angle difference between the inner and outer isophotes 
of the galaxy, and the inclination angle of the disk component will reflect the ellipticity 
of the outer isophotes. If mergers in high-density environments such as galaxy clusters 
induce such changes in ellipticity and position angle with galactocentric radius, then the 
same range of bulge fraction may not select the same galaxies in clusters and in the field. 
In addition to possible deviations from a pure deVaucouleurs law, this behavior of the 
bulge+disk model explains why, for example,  brightest cluster galaxies can often have 
bulge fractions as low as 0.4 \citep{nelson02}. Finally, a single center for the model 
makes it impossible to detect differences in bulge and disk centroids should they manifest 
themselves in some stages of the evolution of galaxies. Unfortunately, the above limitations 
cannot be adequately dealt with by adding more fitting parameters since the basic bulge+
disk model already has enough parameters to make the search for the best-fit solution 
arduous.

It should be kept in mind that the conventional ``bulge/disk'' 
nomenclature does not say anything about the internal kinematics of 
the components.  The presence of a ``disk'' component does not 
necessarily imply the presence of an actual disk since many 
dynamically hot systems also have simple exponential profiles of the 
form given by Equation~\ref{photodisk} below \citep{lin83,kormendy85}.  
Likewise a ``bulge'' may represent a brightened center due to a starburst 
rather than a genuine dynamically hot spheroid.  To avoid any confusion between photometric structures and 
internal dynamics entities, the names ``photobulge'' (for 
``photometric bulge'') and ``photodisk'' (for ``photometric disk'') 
will be used hereafter to refer to the $r^{1/n}$ and exponential 
components of galaxy light profiles respectively.

The 2D galaxy model used by GIM2D has a maximum of twelve parameters: 
the total flux $F$ in data units (DU), the bulge fraction $B/T$ ($\equiv$ 0 for 
pure photodisk systems), the photobulge semi-major axis effective radius $r_e$, 
the photobulge ellipticity $e$ ($e \equiv 1-b/a$, $b \equiv$ semi-minor axis, $a 
\equiv$ semi-major axis), the photobulge position angle of the major axis $\phi_{b}$ 
on the image (clockwise, y-axis $\equiv$ 0), the photodisk semi-major axis 
exponential scale length $r_d$ (also denoted $h$ in the literature), the photodisk 
inclination $i$ (face-on $\equiv$ 0), the photodisk position angle $\phi_d$ on 
the image, the subpixel $dx$ and $dy$ offsets of the model center with respect 
to the thumbnail science image center, the background residual level $db$, and 
the S\'ersic index $n$. Twelve parameters is a maximum since one or more 
parameters can be frozen to their initial values if necessary depending on the scientific 
goals being pursued.  The position angles $\phi_b$ and $\phi_d$ were not forced 
to be equal for two reasons: (1) a large difference between these position angles is 
a signature of barred galaxies, and (2) some observed galaxies do have {\it bona fide} 
photobulges that are not quite aligned with the photodisk position angle. 

Two types of radii are in use in the literature: geometric mean (also known as ``circular'' or ``equivalent'') radii 
and semi-major axis radii. The $same$ name \ie, ``effective'' is used for both types 
of radii, and confusion arises when the type of radius used for a given sample is not 
clearly specified. As noted above, the photobulge effective radius in the GIM2D 
image model is calculated with respect to the semi-major axis of the photobulge 
component. Since the geometric mean (or ``circular'') photobulge effective radius $r_{e,circ}$ is given by 
$\sqrt{ab}$, it is related to the semi-major axis radius $r_{e,sma}$ by the simple 
relation $r_{e,circ}$ = $r_{e,sma} \sqrt{1-e}$. Similarly, the GIM2D semi-major 
axis photodisk scale length $r_{d,sma}$ is related to the circular photodisk scale length by 
$r_{e,circ}$ = $r_{e,sma} \sqrt{1-e_{disk}}$ where $e_{disk}$ is the ellipticity 
of the photodisk. Given the thinness of galaxy disks and the great distances of high-redshift 
galaxies, the GIM2D photodisk model does not need to include a vertical scale height, 
The inclination angle of this infinitely thin photodisk is calculated from the measured 
photodisk ellipticity as $i$ = arccos($\sqrt{1-e_{disk}}$) assuming that face-on 
photodisks are perfectly circular. This relation between photodisk ellipticity and 
inclination is different  from the prescriptions used locally in Tully-Fisher work 
(\citealp[e.g.,][]{courteau97}) which must account for the effect of disk scale height on the 
observed ellipticity of highly-inclined spiral galaxies. 

The first component (``photobulge'') of the 2D surface brightness model used by 
GIM2D to fit galaxy images is a S\'ersic profile of the form:
\begin{equation}
	\Sigma(r) = \Sigma_{e} exp \{-k[(r/r_{e})^{1/n} - 1]\}, 
 	\label{sersic}
\end{equation} 
\noindent where $\Sigma(r)$ is the surface brightness at radius $r$ 
\citep{sersic68}.  The parameter $k$ is set equal to 1.9992$n-$0.3271 
so that $r_{e}$ remains the projected radius enclosing half of the 
light in this component \citep{capaccioli89}.  The classical de 
Vaucouleurs profile has the special value $n$ = 4, and this value was 
chosen for the current analysis.  This choice was motivated by studies 
of bulge profiles in local galaxies.  Locally, there is evidence that 
the bulges of late-type spiral galaxies may be better fitted by an $n$ 
= 1 profile, whereas bright ellipticals and the bulges of early-type 
spiral galaxies follow an $n$ = 4 profile \citep{dejong96, courteau96, 
andredakis98}.  Local late-type galaxies with $n$ = 1 bulges have 
$B/T \leq 0.1$ \citep{dejong96}.  Since such 
bulges contain only 10\% of the total galaxy light, low 
signal-to-noise measurements of late-type high-redshift galaxies make 
it very difficult to determine the S\'ersic index.  On the other hand, 
$n$ is more important for bulge-dominated galaxies, and $n$ = 4 is the 
expected value based on local early-type galaxies.  Knowing that 
bright ellipticals and the bulges of early-type spirals are 
well-fitted by a de Vaucouleurs profile, a $n$ = 4 bulge profile was 
therefore adopted as the canonical bulge fitting model here for the 
sake of continuity across the full range of morphological types.  The 
total flux in the S\'ersic photobulge component is calculated by 
integrating Equation~\ref{sersic} from $r$ = 0 to infinity to obtain:

\begin{equation}
     F_{\rm photobulge} = 2 \pi n e^{k} k^{-2n} r_e^2 \Gamma(2n) \Sigma_{e}
     \label{sersicflux}
\end{equation}

\noindent where $\Gamma$ is the gamma function.  For $n = 
4$, $F_{\rm photobulge} = 7.214 \pi r_{e}^{2} \Sigma_{e}$.  The bulge 
component was collapsed to a point (zero radius) source anytime its 
effective radius was less than 0\arcsecpoint 02 (0.2 WFC pixel) when 
the bulge+disk model fits were performed.

The second component (``photodisk'') of the GIM2D model is a simple 
exponential profile of the form:

\begin{equation}
	\Sigma(r) = \Sigma_{0} exp (-r/r_d). 
	\label{photodisk}
\end{equation}

\noindent $\Sigma_{0}$ is the face-on central surface brightness.  The 
photodisk is assumed to be infinitely thin.  The total flux in the 
photodisk is given by:

\begin{equation}
	F_{\rm photodisk} = 2 \pi r_d^2 \Sigma_{0}.
	\label{photodiskflux}
\end{equation}

\noindent The projected surface brightness distribution of the 
photodisk inclined at any angle $i$ was calculated by integrating 
Equation~\ref{photodisk} over the areas in the face-on photodisk plane 
seen by each projected pixel.  The disk component was collapsed to a 
point (zero radius) source anytime its disk scale length was less than 
0\arcsecpoint 02 (0.2 WFC pixel) when the bulge+disk model fits were 
performed. Equations \ref{sersic} and \ref{photodisk} are given in their 
circularly symmetric form for simplicity, but the GIM2D model certainly does 
not assume circular symmetry since it includes photobulge ellipticity 
and photodisk inclination as fitting parameters.

The optical thickness of disk galaxies remains a hotly debated issue 
locally \citep{valentijn94,giovanelli94,disney92,valentijn90}, and this 
issue obviously has important consequences for the interpretation of local 
and high-redshift photodisk data.  The photodisk optical thickness is not one of the 
fitting parameters, but photodisks can be fitted with surface 
brightness profiles of different optical thicknesses given by the 
parameter $C_{abs}$ in the equation:
 
\begin{equation}
	m_{\rm photodisk, obs} = m_{\rm photodisk, face-on}  + 2.5 \thinspace C_{abs} \thinspace 
{\rm log} (a/b)
	\label{othick-mag}
\end{equation}

\noindent where $m_{\rm photodisk, obs}$ is the observed photodisk 
total magnitude and $m_{\rm photodisk, face-on}$ is the face-on 
photodisk total magnitude.  $a/b$ is again the axial ratio of the 
photodisk component.  Allowable values for $C_{abs}$ are between 0 
(optically thin photodisks) and 1 (optically thick photodisks).  All 
GSS photodisk models used in this analysis were totally optically thin 
($C_{abs} = 0$).  No internal absorption was applied to photobulges.

A PSF-deconvolved semi-major axis half-light radius $r_{hl}$ was also 
computed for each galaxy by integrating the sum of Equations~\ref{sersic} 
and~\ref{photodisk} out to infinity with the best fitting structural parameters. 
This half-light radius may be in error for galaxies with large differences 
between their photobulge and photodisk position angles.

The WFPC2 detector undersampling was taken into account by generating 
the surface brightness model on an oversampled grid, convolving it 
with the appropriate oversampled PSF, shifting its center 
according to $dx$ and $dy$ and rebinning the result to the detector 
resolution for direct comparison with the observed galaxy image. Letting $dx$ 
and $dy$ be free parameters in the fits also helps compensate for possible errors in the 
initial determinations of the galaxy centroid by SExtractor.

\subsection{Types of Fits}\label{typefits}

Before describing each type of fit, the question of what is being fitted should 
be discussed.  Some studies (\citealp[e.g.,][]{schade96}) ``symmetrized'' galaxy 
images before fitting with a bulge+disk model.  Symmetrization removes any 
asymmetric structures around a chosen image pivot point, i.e., the galaxy image 
centroid.  Symmetrization makes sense since the fitting model is itself symmetric, 
and asymmetries can affect the final values for the best-fitting parameters 
(\citealp[e.g.,][]{marleau98} and Section \ref{asym-fit} of this paper) when these asymmetries 
are very strong.  Other studies (\citealp[e.g.,][]{ratnatunga99}) do not use image 
symmetrization.  Although image symmetrization is an option in GIM2D, it was 
not used for the GSS structural analysis.  To symmetrize or not to symmetrize 
is really a ``philosophical'' choice.  The result of image symmetrization depends 
strongly on the choice of pivot point.  An error in pinpointing the location of the 
true pivot point may result in an artificial change in the size of the object being 
fitted.  The ambiguity on the true location of the galaxy centroid will increase 
with the strength of the asymmetric structure.  So, as symmetrization would 
become more and more useful, the error introduced by a bad pivot point is also prone 
to become larger.  Since symmetrization is not used here, possible effects of 
asymmetric structures on fitting parameters should be kept close to mind.  For 
example, star-forming regions and spiral arms in the disk of galaxy can artificially 
increase the measured disk scale length and decrease the measured bulge fraction 
(Note that the problem can also affect symmetrized images if deviant structures 
were symmetric about the centroid of the galaxy).  Fortunately, residual image 
indices (section~\ref{resindices}) can be used to identify objects for which 
structural parameters may have been seriously compromised by asymmetric 
structures. An improved symmetrization procedure in which the total flux removed 
from the input image is minimized might solve the pivot point problem. The effects 
of asymmetries on the measured structural parameters from non-symmetrized 
fits are studied in detail in Section~\ref{asym-fit}.

GIM2D offers three types of galaxy surface brightness fits: (1) 
two-bandpass, separate fits, (2) two-bandpass, simultaneous fits, and 
(3) single-bandpass, multiple image fits.  The first two types of fits 
were used for the GSS structural analysis, and the third one is 
particularly useful for dithered data such as used for HST/NICMOS. 
Science goals dictate the type of fits that should be used.

For the separate, two-bandpass fits, the GSS $I_{814}$ and $V_{606}$ 
thumbnail images were fitted independently, \ie, no fitting parameter 
was constrained to have the same value in both bandpasses.  The only 
connection between the two bandpasses was the use of the $I_{814}$ 
segmentation thumbnail image for {\it both} $I_{814}$ and $V_{606}$ 
fits.  By comparing parameter values in both bandpasses, this type of 
fit can provide valuable information about color gradients in 
photobulges or in photodisks.  These color gradients would yield 
very different photobulge effective radii or photodisk scale lengths 
in the two bandpasses.  Also, the difference in the location of the 
$I_{814}$ and $V_{606}$ model centroids can be used as another image 
asymmetry estimate.  For example, strong blue asymmetric structures 
possibly arising from star formation should perturb the centroid of 
the $V_{606}$ model more than the $I_{814}$ centroid. The main 
disadvantage of this type of fit is that it does not make maximum use 
of all the information available at a given signal-to-noise ratio. In 
the absence of color gradients, all the information should be used by 
fitting both bandpasses simultaneously as described below. 

For the simultaneous, two-bandpass fits, the GSS $I_{814}$ and 
$V_{606}$ thumbnail images were fitted simultaneously with all but 
three fitting parameters forced to take on the same values in both 
bandpasses.  The three fitting parameters allowed to be different were 
the total flux in the model (now $F_{I}$ and $F_{V}$), the bulge 
fraction (now $(B/T)_{I}$ and $(B/T)_{V}$) and the background residual 
level (now $db_{I}$ and $db_{V}$).  This type of fit is, by its nature, 
blind to color gradients in the structural subcomponents, but, in the 
absence of such gradients, it should yield better photobulge and 
photodisk colors.

The third and last type of fit was not used for the DEEP/GSS analysis, but 
it deserves some discussion for its usefulness in fitting stacks of 
dithered images.  Some WFPC2 and many NICMOS imaging programs consist 
of dithered exposures of the same region of the sky.  The offsets 
between the images can be just a few pixels or they may be a 
significant fraction ($\sim 1/3$) of the detector field of view.  The 
offsets are often non-integer pixel shifts, and flux interpolation 
must be used in these cases to combine the images.  Interpolation 
will affect the noise characteristics of the images and should be 
avoided.  Significant flux errors can also be introduced by interpolating 
undersampled image data (\eg, WF and NICMOS/NIC3 cameras). 
The single-bandpass, multiple image fits use the {\it same} 
model to simultaneously fit multiple dithered images.  GIM2D computes 
an image centroid for each image, and the image centroids are passed 
on to the model generating routine so that image models will be 
shifted accordingly.  The main advantage here is the ability to avoid 
interpolating pixel flux.

All three types of GIM2D fits are performed on all pixels flagged as 
object {\it or} background in the SExtractor segmentation image. 
Object areas in the segmentation image are sharply delineated by the 
location of the isophote corresponding to the detection threshold 
(1.5-$\sigma_{bkg}$ here) since SExtractor considers all pixels below 
this threshold to be background pixels. However, precious information 
on the outer parts of the galaxy profile may be contained in the pixels 
below that threshold, and fits should therefore not be restricted only to 
object pixels to avoid throwing that information away. Pixels belonging to 
objects in the neighborhood of the primary object being fitted are masked 
out of the fitting area using the SExtractor segmentation image. The flux 
from the primary object that would have been in those masked areas in the 
absence of neighbors is nonetheless properly included in the magnitude 
measurements given in this paper because magnitudes were obtained by 
integrating the best-fit models over {\it all} pixels.

\subsection{Initial image moments and Fitting Parameter 
Values}\label{initval}

Even though the SExtractor local background was subtracted from each 
science thumbnail image, GIM2D can compute a residual mean background 
level to correct for any error in SExtractor background estimates.  
GIM2D can also be instructed to compute its own estimate of 
$\sigma_{bkg}$.  To compute background parameters, GIM2D uses all the 
pixels in the science thumbnail image flagged as background pixels 
(flag value of zero) in the SExtractor segmentation image.  GIM2D 
further prunes this sample of background pixels by excluding any 
background pixel that is closer than five pixels from any (primary or 
neighboring) object pixels.  This buffer zone ensures that the flux from 
all SExtracted objects in the image below all the 1.5-$\sigma_{bkg}$ 
isophotes does not significantly bias the mean background level 
upwards and artificially inflate $\sigma_{bkg}$.  The GIM2D background 
parameters are tested in Section~\ref{simulations}.  For the GSS fits, 
background parameters were re-calculated with GIM2D before fitting, 
and the residual background levels $db$ were then frozen to their
recalculated values in the surface brightness fits. 

SExtractor object centroid coordinates can be accepted as is, or GIM2D 
can determine its own intensity-weighted object centroid using a 
multi-threshold, minimum area procedure.  This procedure first identifies 
which of three thresholds (10-$\sigma_{bkg}$, 5-$\sigma_{bkg}$, and 
3-$\sigma_{bkg}$) has enough object pixels, as flagged by the SExtractor 
segmentation image, to meet a certain minimum area requirement.  This 
minimum area was set to 8 for the GSS objects.  GIM2D then computes 
intensity-weighted centroid coordinates at the highest level with 
enough pixels.  If none of the three thresholds provides enough pixels to 
calculate the centroid, then GIM2D simply uses all object pixels to 
compute the centroid.

It is possible to place generous initial limits on certain structural 
parameters (total flux, initial photobulge and photodisk scale radii 
and position angles) based on simple image moments widely in use 
throughout the literature.  For a given object, these moments are 
computed about the image centroid over the object pixels.  These 
moments include:

\begin{mathletters}
	\begin{eqnarray}
		M_{tot} & = & \sum_{i,j} I_{ij}
		\label{itot-eq}
		\\
		M_{xx} & = & {1\over{M_{tot}}}\sum_{i,j} x_{ij}^{2} I_{ij}
		\label{xx-eq}
		\\
		M_{yy} & = & {1\over{M_{tot}}}\sum_{i,j} y_{ij}^{2} I_{ij}
		\label{iyy-eq}
		\\
		M_{xy} & = & {1\over{M_{tot}}}\sum_{i,j} x_{ij} y_{ij} I_{ij}
		\label{ixy-eq}
		\\
		M_{rr} & = & {1\over{M_{tot}}}\sum_{i,j} (\sqrt{x_{ij}^{2} + 
		y_{ij}^{2}}) I_{ij}
		\label{ir-eq}
	\end{eqnarray}
\end{mathletters}

\noindent where the sums are over the total number of pixels belonging to the 
object, $x_{ij}$ and $y_{ij}$ are pixel coordinates with respect to 
the object centroid, and $I_{ij}$ is the background-subtracted pixel flux 
in the $(i,j)$th pixel.  The initial total flux estimate for the image 
model is given by Equation~\ref{itot-eq}, and the maximum limit on 
this total flux is set to twice $M_{tot}$.  The initial values for the 
photobulge effective radius and photodisk scale length are both set to 
the intensity-weighted average radius of Equation~\ref{ir-eq}, and the 
maximum limits placed on these two scale radii are set to twice the 
intensity-weighted average radius $M_{rr}$. The minimum limits on the two scale 
radii are set to zero to allow the model to fit unresolved sources if needed.

The object position angle is another parameter that can be set to 
an initial value given by the image moments above as:

\begin{eqnarray}
	\phi = {1\over{2}} \arctan\left({{2 M_{xy}}\over{M_{yy}-M_{xx}}}\right)
	\label{posang}
\end{eqnarray}

So, both $\phi_{b}$ and $\phi_{d}$ are initially set to $\phi$. No 
minimum/maximum limits are placed on the photobulge and photodisk 
position angles since Equation~\ref{posang} is a circular function. 
In the case of a perfectly circular object ($M_{xx} = M_{yy}$), both 
$\phi_{b}$ and $\phi_{d}$ are set to zero. Both parameters are given 
generous initial Metropolis ``temperatures'' (see next section) of 
60\deg.

\subsection{Metropolis Fitting Algorithm}\label{metro}

The 12-dimensional parameter space can have a very complicated 
topology with local minima at low S/N ratios.  It was therefore 
important to choose an algorithm which did not easily get fooled by 
these local minima.  The Metropolis algorithm \citep{metropolis53,saha94} 
was designed to search for optimal parameter values in a complicated 
topology, and it is widely used in many fields of physics and computer 
science.  Compared to gradient search methods, the Metropolis algorithm 
is not efficient, \ie,  it is CPU intensive.  On the other hand, gradient searches 
are ``greedy.''  They will start from initial parameter values, dive in the first 
minimum they encounter and claim it is the global one.

The first step in the fitting process is the ``initial condition filter'' (ICF).  
The ICF coarsely samples the very large volume of structural parameter 
space $V_{0}$ defined by broad limits from the initial image moments 
(Section~\ref{initval}) and user-specified parameter constraints to determine 
a promising sub-volume to be explored by the Metropolis algorithm. The 
uniform ICF sampling for the $i^{th}$ structural parameter follows the 
equation:

\begin{eqnarray}
	\Delta x_{i} = x_{i,0} + (u - {1\over{2}}) T_{i}
	\label{icf-samp}
\end{eqnarray}

\noindent where $\Delta x_{i}$ is a trial step in parameter space, 
$x_{i,0}$ is the $i^{th}$ coordinate of the ICF sampling origin given 
by the image moments and user constraints, $u$ is a uniform deviate 
between 0 and 1, and $T_{i}$ is the ``Metropolis temperature'' of the 
parameter.  The Metropolis temperature of a parameter is given in the
same units as that parameter. The ICF samples parameter space with 
$N_{ICF}$ (set to 400 here) different galaxy image models and keeps 
track of which model yields the highest likelihood.  The appropriate value 
of $N_{ICF}$ depends on how well constrained the search volume is 
initially.  After completion, the ICF sets the sampling origin to the 
parameters of that best ICF model.  The volume $V_0$ is reduced by 
a factor of $N_{ICF}$ by cooling each Metropolis parameter temperature 
according to $T_{i}^{\prime} = T_{i}/(N_{ICF})^{1/n}$ with $n = 
12$ being the number of fitting parameters in the galaxy image model 
used here.  The new sampling origin, the reduced search volume, and 
the cooler Metropolis parameter temperatures are then sent to the 
Metropolis algorithm to continue the search further.

The Metropolis algorithm in GIM2D starts from the sampling origin 
given by the ICF and computes the likelihood $P(\overline{{\bf 
w}}|D,M)$ that the parameter set $\overline{{\bf w}}$ is the true one 
given the data $D$ and the model $M$.  It then generates random 
perturbations $\Delta\overline{\bf x}$ about that initial location 
with a given ``temperature.''  When the search is ``hot,'' large 
perturbations are tried.  After each trial perturbation, the 
Metropolis algorithm computes the likelihood value $P_{1}$ at the new 
location, and immediately accepts the trial perturbation if $P_{1}$ is 
greater than the old value $P_{0}$.  However, if $P_{1}$ $<$ $P_{0}$, 
then the Metropolis algorithm will accept the trial perturbation only 
$P_{1}$/$P_{0}$ of the time.  Therefore, the Metropolis algorithm will 
sometimes accept trial perturbations which take it to regions of lower 
likelihood.  This apparently strange behavior is very valuable: if the 
Metropolis algorithm finds a minimum, it will try to get out of it, 
but it will only have a finite probability (related to the depth of 
the minimum) of succeeding.  The ``temperature'' is regulated 
according to the number of accepted iterations.  If the Metropolis 
accepts too many trial perturbations, then the search is too ``cold,'' 
and the temperature must be increased.  Conversely, if the Metropolis 
rejects too many trial perturbations, then the search is too ``hot,'' 
and the temperature must be decreased.  The Metropolis temperature is 
regulated such that half of the trial perturbations are accepted.  The 
temperature is checked every fiftieth iteration, and the terms of the 
covariance matrix ${\bf s}$ are adjusted according to whether the 
search is too hot or too cold.  The more commonly known ``simulated 
annealing'' technique is a variant and a special case of the 
Metropolis algorithm in which the temperature is only allowed to 
decrease until a  ``ground-state'' is reached.

The step matrix for the trial perturbations $\Delta\overline{{\bf x}}$ 
is given by the simple equation $\Delta\overline{{\bf x}} = {\bf Q} 
\cdot \overline{{\bf u}}$, where the vector $\overline{{\bf u}}$ 
consists of randomly generated numbers between $-$1 and 1.  The matrix 
${\bf Q}$ thus controls the step distribution, and random steps with 
any desired covariance can be generated by solving the equation ${\bf s} = {\bf Q} 
\cdot {\bf Q^{T}}$ through Choleski inversion.  The matrix ${\bf s}$ 
is the local covariance matrix of accepted iterations 
(\cite{vanderbilt84}).  In short, the Metropolis sampling of parameter space 
shapes itself to the local topology.

Convergence is achieved when the difference between two likelihood 
values separated by 100 iterations is less than 3$\sigma$ of the 
likelihood fluctuations.  After convergence, the Metropolis algorithm 
Monte-Carlo samples the region where the likelihood is thus maximized 
and stores the accepted parameter sets as it goes along to build the 
distribution $P(\overline{{\bf w}}|D,M)$.  Once the region has been 
sufficiently sampled ($N_{sample} = 300$ here), the Metropolis 
algorithm computes the median of $P(\overline{{\bf w}}|D,M)$ for each 
model parameter as well as the 68\% confidence limits.  The output of 
the fitting process consists of a PSF-convolved model image $O$, a 
residual image $R$ and a log file containing all Metropolis algorithm 
iterations, the final parameter values and their confidence intervals.

GIM2D creates two output images for each fitted object: an image of 
the PSF-convolved model and an image showing the residuals from the 
bulge+disk model subtraction.  Figure~\ref{gss-image-mosaics} shows 
mosaics (science, GIM2D output models, and residual) of 35 GSS 
galaxies with $I_{814} \leq 24$ chosen at random from the whole GSS 
sample.  All three mosaics use the same greyscale levels.  The 
$I_{814}$ magnitude, bulge fraction and semi-major axis half-light 
radius are given on the original images.

\subsection{Residual Image and Asymmetry Indices}\label{resindices}

GIM2D computes six image indices from the thumbnail residual image 
that can be used to globally characterize the structures left after 
the best galaxy image model has been subtracted. 

The first indices, $R_{T}$ and $R_{A}$, were first applied to distant 
galaxies by \citet{schade95}.  These indices are based on local 
studies of spiral arm patterns \citep{elmegreen92}, and they provide 
an estimate of the overall smoothness of the galaxy image with respect 
to the fitting model.  Following \citet{schade95}, $R_{T}$ and 
$R_{A}$ are defined as:

\begin{mathletters}
	\begin{eqnarray}
		R_T & = & (R_T)_{raw} - (R_T)_{bkg} \nonumber \\
		& = & {{\displaystyle\sum_{i,j} {1\over{2}}| R_{ij} + 
		R_{ij}^{180}|} \over{\displaystyle\sum_{i,j} I_{ij}}} - {{\displaystyle\sum_{i,j} 
		{1\over{2}}| B_{ij} + 
		B_{ij}^{180}|} \over{\displaystyle\sum_{i,j} I_{ij}}}
		\label{rt}
		\\
		R_A & = & (R_A)_{raw} - (R_A)_{bkg} \nonumber \\
		& = & {{\displaystyle\sum_{i,j} {1\over{2}}| R_{ij} - 
		R_{ij}^{180}|} \over{\displaystyle\sum_{i,j} I_{ij}}} - {{\displaystyle\sum_{i,j} 
		{1\over{2}}| B_{ij} - 
		B_{ij}^{180}|} \over{\displaystyle\sum_{i,j} I_{ij}}}
		\label{ra}
	\end{eqnarray}	
\end{mathletters}

\noindent where the $R_{i,j}$'s are pixel values in the residual 
image, and the $R_{ij}^{180}$'s are pixel values in the residual image 
rotated by 180\deg.  Since these raw values $(R_T)_{raw}$ and 
$(R_A)_{raw}$ involve taking absolute values of pixel fluxes, they 
will yield a positive signal even in the sole presence of white noise.  
These raw values must therefore be background-corrected.  The 
background corrections, $(R_T)_{bkg}$ and $(R_A)_{bkg}$, are computed 
over pixels flagged as background pixels in the SExtractor 
segmentation image.  The $B_{i,j}$'s are background pixel values in 
the residual image, and the $B_{ij}^{180}$'s are background pixel 
values in the residual image rotated by 180\deg.  The background 
corrections are computed over a background pixel area equal to the 
pixel areas over which the raw indices were computed. $B_{i,j}$'s 
were randomly drawn from the full pool of all background pixels in the 
thumbnail science image to decrease the vulnerability of background 
parameter determination to possible localized background image artifacts.

There is a fundamental difference between the $R_T$ and $R_A$ indices of 
\citet{schade95} and the ones implemented in GIM2D. Schade et al.  
calculated residual indices within a {\it physical} radius of 5 kpc 
irrespective of the physical size of the whole galaxy.  On the other 
hand, GIM2D computes its $R_T$ and $R_A$ indices within ten circular 
apertures whose radii are multiples  of the galaxy half-light radius 
$r_{hl}$ ranging from 1 $r_{hl}$ to 10 $r_{hl}$.  The GIM2D 
$R_T$ and $R_A$ indices therefore sample the same fractional area for all 
galaxies.  Schade et al.  define normal galaxies as galaxies with 
$R_T+R_A \leq 0.14$.  \citet{im01} were able to define a sample 
of high-redshift E/SO purely based on quantitative morphology with 
only two simple selection criteria ($(B/T)_{I} > 0.4$ and $R_T+R_A 
\leq 0.08$), and \citet{mcintosh01} used $R_T+R_A$ to study the S0 
populations of a sample of local Abell clusters.  Measuring $R_T$ and 
$R_A$ in different bandpasses could also show the wavelength-dependence 
of the strength of residual structures.  If $R_T$ and $R_A$ have larger 
values in bluer bandpasses (\eg, $V_{606}$ versus $I_{814}$), then 
this would suggest that asymmetries arise from star-forming regions. For 
example, there appears to be a correlation between the strength of residual 
structures and star formation rates measured from [OII] emission lines in 
galaxies in poor groups \citep{tran01}.

The next two image indices are from the automated classification 
proposed by \citet{abraham94,abraham96}.  This classification relies on two 
parameters: one measuring the concentration of galaxy light ($C$) and 
the other one measuring image asymmetry ($A$).  The so-called $C-A$ system 
explicitly takes the ellipticity of galaxy images into account instead 
of simply using circular apertures.  The image moments are used to 
define an equivalent elliptical distribution. For each object, the area 
$A_{2\sigma}$ of the 2-$\sigma_{bkg}$ isophote is first computed.  Then, 
following \citet{abraham94}, the definitions of the concentration 
index $C$ and normalized radius $r_{ij}$ are:

\begin{mathletters}
\begin{eqnarray}
	C(\alpha) & = & {{\displaystyle\sum_{(i,j) \in E(r_{ij} \leq \alpha)} I_{ij}}\over{\displaystyle\sum_{(i,j) \in 
	E(r_{ij} \leq 1)} I_{ij}}}
	\label{ca-C}
	\\
	r_{ij} & = & M_{yy} x_{ij}^{2} + M_{xx} y_{ij}^{2} - 2 M_{xy}x_{ij}y_{ij}
	\label{ca-alpha}
\end{eqnarray}
\end{mathletters}

\noindent where $E(r_{ij})$ is an elliptical aperture bounded by 
$r_{ij}$, \ie, $r_{ij}$ is constant on elliptical isophotes, and the 
image moments are normalized such $E(1) = A_{2\sigma}$.  Abraham et 
al.  adopted a value of $\alpha = 0.3$.  $C(\alpha)$ was computed for 
four values of $\alpha$ (0.1, 0.2, 0.3, and 0.4) for the GSS galaxies.  
The $A$ parameter measures image asymmetry in a similar way to $R_A$.  
However, $A$ is computed directly on the science image whereas $R_A$ 
is computed on the residual image from the bulge+disk fit.  $A$ is 
also computed over the area $A_{2\sigma}$, and again similarly to 
$R_A$, a background correction must be applied to raw $A$ values to 
remove the extra positive signal from the background image noise.  The 
$C-A$ indices have been implemented in GIM2D exactly as prescribed by 
\citet{abraham96}.  Subsequent works (\citealp[e.g.,][]{conselice00,wu99}) 
have since shown that choosing an image pivot point which minimizes the 
measured asymmetry greatly improves the classification results, but these 
improvements have not been included in GIM2D yet.

In addition to the previous indices, two new indices, $A_{z}$ and 
$D_{z}$, were defined to quantify residual asymmetric structures.  The 
$A_z$ index is calculated over the pixels belonging to the object or 
the background.  Each pixel is compared to its symmetrical counterpart 
about the center of the object.  If the flux in that pixel is 
$n\sigma_{bkg}$ higher than the flux in its symmetrical counterpart, 
then the pixel is taken to be part of an asymmetrical component.  
$A_{z}$ is the sum of the fluxes of all such pixels normalized by the 
total object flux.  The $A_{z}$ index is computed for $n = $ 2, 3 and 
5 within circular annuli with radii between 1$r_{hl}$ and 10$r_{hl}$.  
A statistical background correction similar to the ones used for 
$(R_T)$ and $(R_A)$ was applied to the raw $A_z$ values. The $D_z$ 
index takes advantage of the isophotal shape of the object as measured 
by SExtractor.  $D_z$ is the sum of the fluxes of the object pixels (as 
given by the segmentation image) with symmetrical counterparts about 
the object center which do not belong to the object.  This sum is normalized 
by the total object flux.  $D_{z}$ is computed {\it outside} of one half-light radius, 
and it is sensitive to asymmetries such as tidal arms.

\section{Structural Catalogs}\label{catalogs}

The images of GSS galaxies were fitted with both the separate and 
simultaneous fitting procedures, and extensive sets of GSS simulations 
were performed for both (Section~\ref{simulations}).  As a result, 
four structural parameter catalogs are presented in this paper: two 
science catalogs and two simulation catalogs.  These catalogs are 
automatically generated from the GIM2D output log files by a set of 
scripts.  These scripts calculate all the final parameter values including 
physical lengths and rest-frame quantities (Section~\ref{restcat} 
below) with full Monte-Carlo propagation of the parameter errors 
(Section~\ref{errorbars} below). The Keck/LRIS spectroscopic 
redshifts used to compute physical radii in kiloparsecs and rest-frame 
magnitudes and colors for the whole galaxy, the photobulge and the 
photodisk  are taken from \citet{phillips02}. The observational, 
reduction and analysis procedures used to measure these redshifts 
are fully described in their paper.

\subsection{Description}\label{catdesc}
The contents of the science catalogs are listed in 
machine-readable Tables~\ref{sepfit-cat-desc} (separate fits) 
and~\ref{simulfit-cat-desc} (simultaneous fits).  The separate fit 
science catalog contains 7450 objects, and each object has 330 columns 
of parameter information (including error bar columns).  Simultaneous 
fits were performed only to objects with Keck/LRIS redshifts from the 
DEEP survey of the Groth Strip.  The simultaneous catalog thus contains 
648 objects with 279 columns of information for each one, including error 
bar columns.  There are two error columns associated with 
each entry described in Tables~\ref{sepfit-cat-desc} 
and~\ref{simulfit-cat-desc}: one for the lower 68$\%$ confidence bound 
and the other for the upper 68$\%$ bound.  Most of the column descriptions in 
Tables~\ref{sepfit-cat-desc} and ~\ref{simulfit-cat-desc} are 
self-explanatory, but some of them require further details:

{\it DEEP/GSS IDs ``gssid'':} The internal DEEP/GSS object IDs (``{\it 
gssid}'') are given by the format FFC-XXYY where FF is the GSS field 
(Table~\ref{hst-images-lst}, Column 1), C is the WFPC2 chip number , 
and XX and YY are the object coordinates on the chips in units of 10 
pixels.  These internal IDs are extended with letters (``a'', ``b'', ``c'', etc.)  
when a group of objects are close enough together that they would all 
have the same primary ID. In addition to these internal DEEP/GSS 
objects ID's, the catalogs list the J2000.0 coordinates of each object.

{\it Physical scale lengths:} All angular scale lengths ($r_{hl}$, 
$r_e$, and $r_d$) were converted to physical lengths according to the 
equation:

\begin{eqnarray}
R = r {c\over{1000H_0}} {1\over{1+z}}\int_{0}^{z} {dz^{\prime}\over{\sqrt{\Omega_m (1+z^{\prime})^3 + \Omega_\Lambda}}}
\label{ang-diam}
\end{eqnarray}

\noindent where $z$ is the redshift, $c$ is the speed of light, $r$ is the measured angular 
scale length in radians, and $R$ is the corresponding physical length 
in kiloparsecs \citep{hogg99}.  Equation~\ref{ang-diam} is only valid for 
flat ($\Omega_k \equiv  1 - \Omega_m - \Omega_\Lambda = 0$) cosmologies.

{\it Position angles on sky:} The GIM2D photobulge and photodisk position angles 
$\phi_{b,image}$ and $\phi_{d,image}$ measured on the 
images clockwise with respect to the positive $y$-axis were converted 
to real position angles $\phi_{b,sky}$ and $\phi_{d,sky}$ on the sky 
using the telescope position angles stored in the HST image headers.

{\it AB Magnitudes:} Even though Vega-based magnitudes are primarily 
used in this paper, Tables~\ref{sepfit-cat-desc} and~\ref{simulfit-cat-desc} 
also provide AB magnitudes.  To go from Vega to AB magnitudes, one 
uses $I_{814} ({\rm AB}) = I_{814} ({\rm Vega}) + 0.44$ and $V_{606} 
({\rm AB}) = V_{606} ({\rm Vega}) + 0.11$.

{\it Galaxy rest-frame $B$-band photobulge fraction:} The rest-frame photobulge 
fraction is given by the simple equation:

\begin{eqnarray}
	(B/T)_{B, rest} = 10.0^{(M_{B, {\rm galaxy}}-M_{B, {\rm	bulge}})/2.5}
	\label{restbt}
\end{eqnarray}

\noindent where $M_{B, {\rm galaxy}}$ and $M_{B, {\rm bulge}}$ are the 
rest-frame $B$-band magnitudes of the galaxy and photobulge respectively.  
Since different $k$-corrections apply to the photobulge and photodisk stellar 
populations, the observed bulge fraction of a galaxy will change with 
redshift.  So, rest-frame photobulge fractions make more uniform photobulge 
fraction selections possible.

\subsection{Rest-Frame Quantities}\label{restcat}

Rest-frame quantities in the GIM2D/GSS structural catalogs were 
calculated using two very different sets of $k$-corrections.  These 
two sets provide independent checks of the reliability of the 
resulting rest-frame quantities.  Rest-frame quantities were 
calculated independently for the total galaxy, the photobulge and the 
photodisk.

The first set of $k$-corrections (referred to as Gronwall 
$k$-corrections throughout this paper) is based on the work of 
\citet{gronwall95}.  The corrections are based on eleven 
theoretical galaxy spectral energy distributions from the 1995 Bruzual 
and Charlot models.  The parameters of these theoretical SEDs are 
given in \citet{gronwall95}.  Some of these SEDs include dust and 
star-bursting populations.  The rest-frame magnitudes and colors of 
any galaxy are obtained by interpolating the SEDs.  The input 
quantities are galaxy redshift, Vega-based $I_{814}$ magnitude and 
$V_{606}-I_{814}$ color.  The output consists of rest-frame ${BVRI}$ 
absolute magnitudes, rest-frame $(U-B)$ and $(B-V)$ colors, and 
rest-frame $B$ and $K$ magnitudes.

The second set of $k$-corrections (referred to as Willmer-Gebhardt or 
WG $k$-corrections throughout this paper, \citealp{gebhardt02}) is 
based on actual galaxy spectra.  These spectra are taken from the 
Database of UV-Optical spectra of Nearby Quiescent and Active Galaxies 
\citep{kinney96,schmitt97}.  This database has recently been expanded to 
include 99 galaxies, 48 of which have full wavelength coverage from 
1200 to 10000\AA\thinspace\thinspace with a combination of International Ultraviolet 
Explorer (IUE) and ground-based spectra.  Filter bandpasses are 
convolved with the galaxy spectra to produce rest-frame $(U-B)$ colors 
and $B$-band $k$-corrections $k_B$.  A polynomial is fitted to the 
observed $(V_{606}-I_{814}$) colors and the rest-frame $(U-B)$ SED 
colors in each redshift range.  The best-fit polynomial reproducing 
the data over the redshift range $0.1-1.1$ is given by:

\begin{eqnarray}
(U-B)_{WG} & = &-0.8079-0.049752z-1.6232z^{2}+1.04067z^{3} \nonumber
\\
& & +1.5294z^{4} -0.41190z^{5}-0.56986z^{6}+(0.61591\nonumber
\\
& & +1.07249z-2.2925z^{2}+1.3370z^{3})(V_{606}-I_{814})\nonumber
\\
& & +(0.280481-0.387205z+0.043121z^{2}) \nonumber
\\
& & (V_{606}-I_{814})^{2}
\label{ub-wg}
\end{eqnarray}

Similarly, the $B$-band $k$-corrections are given by:

\begin{eqnarray}
k_{BI} & = & 0.0496+0.46057z+1.40430z^{2}-0.19436z^{3}\nonumber
\\
& & -0.2232z^{4}-0.36506z^{5}+0.17594z^{6}+(2.0532\nonumber
\\
& & -2.8326z+1.05580z^{2}-0.67625z^{3})(V_{606}-I_{814}) \nonumber
\\
& & +(0.10826-0.68097z+0.61781z^{2})(V_{606}-I_{814})^{2},\nonumber
\\
\label{kb-wg}
\end{eqnarray}

\noindent and the rest-frame $B$-band magnitude in the 
Willmer-Gebhardt system is given by:

\begin{eqnarray}
	M_{B, {\rm WG}} = I_{814} - DM(\Omega_m, \Omega_{\Lambda}, \Omega_k)  + 
	k_{BI}
\label{MB-wg}
\end{eqnarray}

\noindent where $DM(\Omega_m, \Omega_{\Lambda}, \Omega_k)$ is the 
distance modulus for the adopted cosmology. See \citet{gebhardt02} for more details.

\subsection{Error Estimates}\label{errorbars}

All parameter error estimates in the GIM2D/GSS structural parameter 
catalogs are $68\%$ confidence limits.  Many of these error estimates are 
asymmetric since they were derived through full Monte-Carlo 
propagations of the parameter probability distributions 
$P(\overline{{\bf w}}|D,M)$ computed by GIM2D through all the 
transformations required to calculate a given final parameter.  This 
process takes into account all the Gaussian and non-Gaussian 
covariances among the parameters.  To illustrate the process, consider 
the observed photobulge $I_{814}$ magnitude of a galaxy.  This 
quantity depends on both the total galaxy model flux and the observed 
$F814W$ bulge fraction.  So, the parameter probability distributions 
$P(F_{814}|D,M)$ and $P(B/T|D,M)$ were first Monte-Carlo sampled 500 
times, and a photobulge magnitude was calculated each time using 
Equation~\ref{magdefI}.  The resulting 500 photobulge magnitudes were 
then sorted, and the lower and upper $68\%$ confidence error estimates were 
derived from that sorted distribution.  Going one step further, the 
$V_{606}$ photobulge magnitude can be computed in exactly the same way 
as $I_{814}$, and the resulting 500 $V_{606}-I_{814}$ colors can then 
be transformed to photobulge rest-frame $B$-band magnitudes using 
Equations~\ref{kb-wg} and~\ref{MB-wg} so that one can in turn compute 
the $68\%$ lower and upper error estimates on the photobulge absolute 
magnitude.

\section{Simulations}\label{simulations}

Three sets of GIM2D simulations were run to characterize the 
systematic biases and random errors in the GIM2D/GSS structural 
measurements.  These simulations are a key element in the 
interpretation of the observations, and the simulation catalogs are 
presented in the same way as real catalogs to emphasize this point.  
Given any real science plot, it is straightforward to select sets of 
simulated galaxies with the same selection criteria as the 
observations to immediately evaluate the biases and errors present in 
the data shown on that plot.  The first set of simulations applies to 
separate fits of $I_{814}$ and $V_{606}$ GSS galaxy images (Section~\ref{sepfit}).  
It contains 5995 simulations, comparable in size to the real science 
catalog.  The second set includes 5195 GSS simultaneous 
$V_{606}$/$I_{814}$ fit simulations, and these simulations cover the 
full range of observed GSS photobulge and photodisk $V_{606}-I_{814}$ 
colors (Section~\ref{simulfit}).  The two sets of simulations above 
only include smooth galaxy image simulations. The effects of 
asymmetric structures on the measured structural parameters were 
explored with a third set of simulations described in 
Section~\ref{asym-fit}.

\subsection{Separate Fits}\label{sepfit}
For the GSS separate fit simulations, 5995 smooth galaxy image models 
were created with structural parameters uniformly generated at random 
in the following ranges: $20.0 \leq I_{814} \leq 25.0$, $0.0 \leq B/T 
\leq 1.0$, $0 \leq r_{e} \leq 0\arcsecpoint 7$, $0.0 \leq e \leq 0.7$, 
$0 \leq r_{d} \leq 0\arcsecpoint 7$, and $0 \leq i \leq 85\deg$.  The 
S\'ersic photobulge index was held fixed at $n = 4$ for all models.  
Both photobulge and photodisk position angles were fixed to 
90\deg\thinspace\thinspace for all simulations, and the bulge and 
disk sizes were uniformly generated in the log of the size ranges above.  
The goal of these simulations is to characterize biases and errors and 
not to simulate what the real Universe would look like through the 
GIM2D observational ``lens.''  The uniformity of the parameter 
distributions adopted here is therefore perfectly suitable to the task 
even though real galaxy parameters (\eg, bulge fraction) may not be 
so distributed.  In the same spirit, no correlations were imposed 
between the input parameters despite the fact that some parameters 
(\eg, $r_e$ and $r_d$) may be correlated in some types of galaxies 
(\citealp[e.g.,][]{courteau96}).

Each simulation was convolved with a $F814W$ TinyTim PSF. This PSF had 
the same parameters as the TinyTim $F814W$ PSFs used in the GSS 
analysis (Section~\ref{psf}).  The same PSF was used in both creating 
and analyzing the simulations, so the results will not include any 
error in the structural parameters due to PSF mismatch.  Poisson 
deviates were used to add photon noise due to galaxy flux into the 
simulations.  The noisy images were then embedded in a 20\arcsec 
$\times$ 20\arcsec section of one of the real $F814W$ GSS images to 
provide a real background for the simulations.  In addition to sky 
photon noise and detector read-out noise, the real background noise 
includes brightness fluctuations of very faint galaxies below the 
detection threshold.  The simulations were SExtracted with exactly the 
same SExtractor parameter files (Sections~\ref{sexdetect} 
and~\ref{deblend}) as used for the GSS analysis, and GIM2D extracted 
science and segmentation thumbnails from the simulations following 
exactly the same steps as for the real galaxies 
(Section~\ref{thumbim}).  Finally, the GIM2D output log files were 
processed through the same scripts to produce a catalog of final 
recovered structural parameters.  The content of this catalog is 
listed in Table~\ref{sep-simcat}.

\subsubsection{Systematic and Random Errors}\label{sep-simerrors}

For the sake of simplicity, the main tool adopted here to visualize 
errors is a set of two-dimensional maps giving systematic and random 
errors at each position.  It should therefore be kept in mind that 
these maps can only offer a limited representation of the complex 
multidimensional error functions.  As the large number of parameters 
in Tables~\ref{sep-simcat} and~\ref{simul-simcat} indicates, a full 
description of all systematic and random errors over all of bulge+disk 
multivariate structural space would considerably add to the length of 
this paper.  Therefore, the error analysis presented in this section 
will focus on only three main galaxy structural parameters: total 
apparent magnitude, bulge fraction and half-light radius.  Errors on 
any other set of parameters can be described in the same way, and the 
simulation catalog can be used to tailor error analyses to the needs 
of the specific science goals being pursued.

The error maps can be cast in terms of input or recovered coordinates, 
and the choice of coordinate system depends on how the error maps will 
be used.  Input coordinates (\ie, the ``true'' coordinates) can be used to 
compute errors that are to be applied to theoretical galaxy structural 
catalogs in order to convert them to observed quantities (\citealp[e.g.,][]{simard02}.  
To illustrate this process, let $\overline{w_T}$ be the position of a mock 
galaxy in theoretical structural space, and let $r_{hl,T}$ be its theoretical 
half-light radius.  If the simulation catalog shows that the recovered half-light 
radii of galaxies at $\overline{w_T}$ are systematically in error by an amount $\Delta 
r_{1}$, then let $r^{\prime}_{hl,T} = r_{hl,T} + \Delta r_{1}$.  This 
new radius $r^{\prime}_{hl,T}$ is not yet the same as an observed radius 
as it does not include a random error.  The random error on the 
half-light radius $\sigma(r_{hl,T})$ at $\overline{w_T}$ can also be 
calculated from the simulation catalog, and another radius correction 
$\Delta r_{2}$ drawn at random from a Gaussian distribution of width 
$\sigma(r_{hl,T})$ can be applied to $r^{\prime}_{hl,T}$ to produce 
the final ``observed'' theoretical half-light radius.  The second set 
of coordinates, the recovered quantities, is the simulation equivalent 
of observed quantities, and error maps cast in those coordinates can 
be directly compared to the real data to see how important errors are 
in different regions of the observational space.

Figures~\ref{err-map-i814-sepf},~\ref{err-map-rhl-sepf}, and 
~\ref{err-map-bt-sepf} show maps of errors on the galaxy total 
magnitude $I_{814}$, galaxy half-light radius $r_{hl}$ and galaxy 
bulge fraction $(B/T)$ respectively as a function of galaxy magnitude 
and galaxy half-light radius for the DEEP/GSS separate structural 
fits.  The two top panels in each figure show the mean parameter error 
(left-hand panels, top number in cells) and the 1-$\sigma$ parameter 
random error (right-hand panels, top number in cells) as a function of 
input galaxy magnitude and size.  Each cell also gives the number of 
simulations created for that cell (bottom number).  The simulations 
are not evenly distributed over the galaxy magnitude-log size plane 
since the simulations were uniformly generated in log $r_{e}$ and in 
log $r_{d}$.  The lower left-hand panels show the mean parameter 
error as a function of recovered galaxy magnitude and size, and the lower 
right-hand panels of the three figures show the actual DEEP/GSS 
magnitude-size data.  The unresolved objects (log $r_{hl,obs} \le 
-1.5$) are nicely separated from the galaxies in these panels.

Figure~\ref{err-map-i814-sepf} shows that the systematic errors on 
$I_{814}$ galaxy magnitudes start to become significant ($\Delta 
I_{814} \simeq -0.2$) fainter than $I_{814} = 23.5$, and that, at a 
given magnitude, errors are larger for the largest galaxies in the 
simulations (log $r_{hl,input} \ge -0.25$).  The random magnitude 
errors are about 0.05 for $I_{814} \le 23.0$ and 0.13 for $I_{814} > 
23.0$.  The systematic errors cast in terms of input or recovered 
coordinates are essentially the same since the errors are not large 
enough compared to the cell sizes (0.5 mag and 0.5 log $r_{hl}$) to 
shift simulations from cell to cell.  If galaxies were pure bulges or 
pure disks, then according to equations~\ref{sersicflux} 
or~\ref{photodiskflux}, systematic magnitude and size errors should be 
anti-correlated, \ie, given an observed surface brightness profile 
and an underestimate, say, of the total flux (positive magnitude 
error), the profile modelling should try to compensate for that 
magnitude underestimate by introducing a negative error in the size.  
This anti-correlation should still hold for composite systems.  A 
comparison of Figures~\ref{err-map-i814-sepf} 
and~\ref{err-map-rhl-sepf} does show that magnitude and size errors 
are typically anti-correlated.  Magnitude and size errors are 
important for fainter and larger galaxies since their relatively low 
surface brightness makes them more vulnerable to sky estimate errors.

The error maps for the galaxy bulge fraction 
(Figure~\ref{err-map-bt-sepf}) show that bulge fractions are 
underestimated by about 0.15 at magnitudes fainter $I_{814} = 23.5$ 
with random errors around 0.25.  However, Figure~\ref{err-map-bt-sepf} 
is not really the best way to truly understand the behavior of the 
recovered bulge fractions.  There are in fact two {\it expected} 
biases in the bulge fractions, and these biases arise from two 
ingredients of the DEEP/GSS bulge+disk analysis: (1) bulge fractions 
are constrained to stay between 0 and 1, and (2) bulge+disk models 
were fitted to {\it all} detected objects irrespective of the 
signal-to-noise ratio (S/N) of their images.  The constraint on the 
bulge fraction forces the recovered bulge fractions of both very low 
($B/T \simeq 0$) and very high ($B/T \simeq 1$) systems to scatter 
above zero and below one, and this bias will affect all galaxies 
irrespective of their S/N ratios.  The second bias is inherent to 
bulge+disk model fits to objects with different S/N ratio.  Previous 
studies have adopted a two-tier approach to this problem.  Schade et 
al.  (1995, 1996) first fit pure bulge or pure disks to their objects 
and then decide upon visual inspection of the residuals whether a 
bulge+disk model would be more appropriate.  Ratnatunga et al.  1999 
fit bulge+disk models to objects above a certain signal-to-noise, and 
objects below that threshold are fitted only with either a pure bulge 
or a pure disk model.

A different approach was taken here to deal with the bulge+disk S/N 
bias.  Bulge+disk models were fitted to all detected objects here for 
simplicity and for the sake of producing homogeneous structural 
catalogs.  Bulge+disk models will converge to a pure bulge or a pure 
disk model only when the signal-to-noise ratio is high enough to 
definitely establish the presence of one and only one structural 
component.  At low S/N ratios, the model will always be able to 
``slip'' in both structural components.  For example, the model could 
make use of a very large disk component to compensate for an 
underestimate of the sky level that may have been computed during the 
fit to a pure bulge system, and this would artificially decrease the 
recovered bulge fraction.  Low signal-to-noise can also be responsible 
for hiding the outer wing of steep surface brightness profiles such 
as the r$^{1/4}$ profile into the background noise and thus making them 
harder to identify.  Figure~\ref{btm-btbias-sepf} best shows the bulge 
fraction biases for the DEEP/GSS separate structural fits.  
Figure~\ref{btm-btbias-sepf} is similar to previous error maps except 
that the errors are now computed over input bulge fraction and input 
galaxy magnitude instead of magnitude and size.  As expected, the 
$B/T$ systematic errors in Figure~\ref{btm-btbias-sepf} show that (1) 
$B/T$ is indeed overestimated in the first $B/T$ bin, (2) $B/T$ is 
underestimated in the last $B/T$ bin, and (3) the magnitudes of the 
discrepancies increase with magnitude.  The homogeneous approach to 
bulge+disk model fitting adopted here is valid as long as the results 
are used in conjunction with careful error characterization from the 
simulation catalogs.

\subsection{Simultaneous Fits}\label{simulfit}

The GSS simultaneous fit simulations use the $I_{814}$ simulations of 
Section~\ref{sepfit} as a starting point.  A companion $V_{606}$ 
simulation was created for each $I_{814}$ image with the same 
structural parameters except for total flux and bulge fraction.  The 
$V_{606}$ total flux and bulge fraction were calculated from the 
$F814W$ total flux and bulge fraction and from randomly generated 
photobulge and photodisk $(V_{606}-I_{814})$ colors.  The photobulge 
and photodisk $(V_{606}-I_{814})$ colors were uniformly and 
independently generated in the range 0.5-2.2.  This range of colors 
spans the full range of observed colors out to a redshift of $z = 1.1$ 
in the DEEP/GSS survey \citep{phillips02}, and it allows one to 
study the effects on fitting results of differences in photobulge and 
photodisk colors.  For $I_{814}$ pure photobulge systems ($(B/T)_{I} = 
1.0$), the $F606W$ bulge fraction and total flux are given by:

\begin{mathletters}
\begin{eqnarray}
(B/T)_{V} & = & 1.0\label{btvbti1}
\\
F_{tot,V} & = & F_{tot,I}{t_{V}\over{t_{I}}}
10^{(1.26-(V-I)_{\rm photobulge})/2.5}\label{ftotvbti1}
\end{eqnarray}
\end{mathletters}

\noindent where $F_{tot,I}$ and $F_{tot,V}$ are total $F814W$ and $F606W $
galaxy model fluxes in DU respectively, $t_{I}$ and $t_{V}$ are the 
$F814W$ and $F606W$ total exposure times (4400 seconds and 2800 seconds), 
and $(B/T)_{I}$ and $(B/T)_{V}$ are the $F814W$ and $F606W$ photobulge 
fractions. The zeropoint difference between $V_{606}$ and $I_{814}$ 
is 1.26.

For $I_{814}$ pure photodisk systems ($(B/T)_{I} = 0.0$), the $F606W$ bulge fraction 
and total flux are given by:

\begin{mathletters}
\begin{eqnarray}
(B/T)_{V} & = & 0.0\label{btvbti0}
\\
F_{tot,V} & = & F_{tot,I}{t_{V}\over{t_{I}}}
10^{(1.26-(V-I)_{\rm photodisk})/2.5}\label{ftotvbti0}
\end{eqnarray}
\end{mathletters}

For $I_{814}$ composite galaxy systems ($0 < (B/T)_{I} < 1$), the 
$F606W$ bulge fraction and total flux are given by the equations:

\begin{mathletters}
\begin{eqnarray}
(B/T)_{V} & = & \left(   {1-(B/T)_{I}\over{(B/T)_{I}}}
10^{\Delta(V-I)/2.5} + 1.0\right)^{-1}
\label{btvc}
\\
F_{tot,V} & = & F_{tot,I}{(B/T)_{I}\over{(B/T)_{V}}}
{t_{V}\over{t_{I}}} 10^{(1.26-(V-I)_{\rm photobulge})/2.5}\nonumber
\\
\label{ftotvc}
\end{eqnarray}
\end{mathletters}

\noindent where $\Delta(V-I) = (V-I)_{\rm photobulge} - (V-I)_{\rm photodisk}$.

After adding in Poisson noise, the $V_{606}$ simulations were also 
embedded in the corresponding 20\arcsec $\times$ 20\arcsec section of one 
of the real $F606W$ GSS images.  This section of the sky was identical 
to the one used for the $I_{814}$ simulations.  As was done for the separate 
fit simulations, the simultaneous fit simulations were processed in 
exactly the same way as the real galaxies to produce a catalog whose 
content is listed in Table~\ref{simul-simcat}.  Note that, as for the 
observations, the $I_{814}$ segmentation thumbnail images were used 
for both bandpasses in the simultaneous fits.

\subsubsection{Systematic and Random Errors}\label{simul-simerrors}

Figures~\ref{err-map-i814-simulf},~\ref{err-map-rhl-simulf}, and~\ref{err-map-bt-simulf} 
show maps of errors on the galaxy total magnitude $I_{814}$, galaxy half-light 
radius $r_{hl}$ and galaxy bulge fraction $(B/T)$ respectively as a function of 
galaxy magnitude and galaxy half-light radius for the DEEP/GSS simultaneous structural 
fits (see Figures~\ref{err-map-i814-sepf},~\ref{err-map-rhl-sepf}, and~\ref{err-map-bt-sepf} 
for the corresponding separate fit results).  The two top panels show the mean parameter error (left-hand 
panels, top number in cells) and the 1-$\sigma$ parameter random error (right-hand panels, 
top number in cells) as a function of input galaxy magnitude and size.  
Each cell also gives the number of simulations created for that cell 
(bottom number).  Again, the simulations are not evenly distributed 
over the galaxy magnitude-log size plane since the simulations were 
uniformly generated in log $r_{e}$ and in log $r_{d}$.  The lower 
left-hand panels show the mean parameter error as a function of 
recovered galaxy magnitude and size.  The lower right-hand panels of 
the three figures show the actual DEEP/GSS magnitude-size data.  The 
number of data points is not nearly as large as in 
Figures~\ref{err-map-i814-sepf},~\ref{err-map-rhl-sepf}, and 
~\ref{err-map-bt-sepf} since simultaneous bandpass structural fits 
were performed only on DEEP/GSS galaxies with secure Keck/LRIS redshifts.  
The errors from the simultaneous structural fits behave the same way 
as the errors from the separate fits, and simultaneous fit errors seem 
to be slightly smaller than those from the separate fits as one would 
expect from simultaneously using all the information content of both 
bandpasses.  However, the improvement in the errors may not be as 
marked as expected since the simultaneous fit simulations included 
varying bulge fraction and colors as additional input parameters.

Bulge and disk colors were included in the simultaneous structural fit 
simulations with the goal of testing how well they are recovered in the 
fits.  Figure~\ref{rec-inp-bulge-disk-colors} shows the structural 
component colors recovered by GIM2D from simultaneous structural fit 
simulations for galaxies with $r_{hl,814} \ge 0\arcsecpoint 15$, and 
$r_{hl,606} \ge 0\arcsecpoint 15$.  The $V_{606}$ limits for the 
galaxy, bulge, and disk magnitudes were set to 26.0.  The nine panels 
show recovered $V_{606}-I_{814}$ colors versus input $V_{606}-I_{814}$ 
colors for the galaxy as a whole (top panels), the bulge (middle panels) and 
the disk (bottom panels) in three different magnitude ranges.  The colors are all 
well recovered by the fits.  The mean and rms color difference in the 
three magnitude ranges are (0.002, 0.020), (0.010, 0.034), and (0.063, 
0.087) for the galaxy as a whole, ($-$0.001, 0.255), (0.022, 0.120), and 
(0.034, 0.186) for the bulge and (0.008, 0.047), (0.006, 0.211), and 
(0.030, 0.304) for the disk. There are no significant systematic color errors, and 
the rms scatter increases with magnitude. 

Although the recovered colors show no systematic offsets from the 
input colors, there are interesting outliers in some panels of 
Figure~\ref{rec-inp-bulge-disk-colors}.  In the leftmost middle panel, 
some recovered bulge colors are much too blue compared to their input 
colors.  The three most discrepant bulges ($(V_{606}-I_{814})_{input} 
\ge 1.9$ and $(V_{606}-I_{814})_{recovered} \le 1.3$) are all very red 
bulges with very blue disks ($(V_{606}-I_{814})_{input,disk} \le 
0.90$), and their effective radius differs from the scale length of 
their disk by a factor of five or more.  In contrast, the 
central middle panel ($21 \le I_{814} (bulge) < 22.0$) shows recovered 
bulge colors which are too red for their input colors.  The two 
outlier bulges ($(V_{606}-I_{814})_{input} \le 1.0)$ with red 
recovered colors ($(V_{606}-I_{814})_{recovered} \ge 1.5$) are quite 
blue compared to their disks ($(V_{606}-I_{814})_{input,disk} \ge 
2.1$).  One of the bulges has an effective radius that differs by a 
factor of 10 from the disk scale length, but the other has an 
effective radius comparable to the disk scale length.  The central 
bottom panel shows recovered disk colors that are too blue compared to 
their input values.  The two rightmost bottom outliers are very red 
disks with bluer bulges ($(V_{606}-I_{814})_{input,bulge} \le 1.0$), and their 
effective radius is different from the disk scale length by a factor 
of 20-23!
 
The outliers in Figure~\ref{rec-inp-bulge-disk-colors} lead to an 
important question: Is there a combination of bulge fraction, 
bulge/disk size ratio $r_{e}/r_{d}$, and bulge/disk colors for which 
bulges can be mistaken for disks and vice versa?  
Figure~\ref{bt-irr-nobias} shows the systematic error (mean error) on 
bulge fraction as a function of input bulge fraction and input 
log bulge/disk size ratio ($r_{e}/r_{d}$) for both the separate 
structural fit (SPF) simulations and the simultaneous structural fit 
(SMF) simulations.  There are no regions of that plane in which bulge 
fractions are systematically in error.  This confirms the absence of 
systematic deviations in bulge and disk colors in 
Figure~\ref{rec-inp-bulge-disk-colors}.  However, there are places 
(especially for very small or very large bulge/disk size ratios) where 
the minimum or the maximum bulge difference is quite large, and these 
extrema can account for the kind of color outliers seen in 
Figure~\ref{rec-inp-bulge-disk-colors}.

\subsection{Effects of Asymmetric Structures On Fitting 
Parameters}\label{asym-fit} 

Non-smooth local features in a galaxy 2D light profile can alter the 
best parameters derived with GIM2D depending on their brightnesses and 
positions in the galaxy.  For example, a very bright feature at the 
center of the galaxy will cause the bulge component to be 
overestimated.  The effects of clumps or asymmetric features on the 
extracted smooth 2D profile parameters were studied by adding an 
asymmetric light component, in the form of one or multiple ``blobs or 
HII regions,'' \ie, unresolved sources convolved with the PSF, to 
simulated smooth 2D profile images.

The input parameters for generating the asymmetric features are the 
number of HII regions $n_{HII}$, the total flux in the HII regions as 
a fraction $f_{HII}$ of the total galaxy flux, and the HII regions' 
maximum galactocentric distance $r_{HII}$ in units of galaxy 
half-light radius.  The positions of the HII regions were randomly 
distributed within a circular aperture defined by $r_{HII}$.  No 
overall ellipticity or radial exponential weight was given to the 
spatial distributions of the HII regions.  A radially weighted 
elliptical HII region distribution could originate with HII regions 
linked with an inclined galaxy disk, but the ``HII regions'' here are 
meant to represent all unresolved asymmetric structures, and 
asymmetric structures may or may not necessarily be associated with 
the disk components of galaxies.

The total asymmetric flux was distributed among HII regions using a 
simple recipe (below) with no attempt to include, say, a realistic HII 
luminosity function.  The simple recipe produced asymmetric structures 
that visually looked reasonable.  According to this adopted recipe, 
the flux $F_{{\rm HII}, i}$ allocated to the $i^{th}$ HII region was 
generated at random between 0 and

\begin{eqnarray}
	F^{max}_{{\rm HII}, i} =  {1\over{n_{HII}-(i-1)}}\left(f_{HII}F_{\rm total, 
	galaxy} - \sum_{j = 0}^{i-1} F_{{\rm HII}, j}\right) \nonumber
	\\
	\label{HII-flux}
\end{eqnarray}

\noindent where $1 \leq i \leq n_{HII}$, $F_{\rm total, galaxy}$ is 
the total model galaxy flux, $F^{max}_{{\rm HII}, i}$ is the maximum 
flux available to the $i^{th}$ HII region, $F_{{\rm HII}, j}$ is the 
flux that was actually allocated to the $j^{th}$ HII region, and 
$F_{{\rm HII}, j = 0} = 0$.  So, the bracket in 
Equation~\ref{HII-flux} contains the total unallocated HII flux that 
remains after $i-1$ regions have been created.  For $i = n_{HII}$, 
$F_{{\rm HII}, i}$ is automatically set to the left-over HII flux.

Asymmetric features superposed on the smooth profile were generated 
randomly for $n_{HII} = 5$ and $r_{HII} = 1.5 r_{hl}$.  Fifteen {\it 
identical} models were created for each simulated galaxy.  The first 
five models had no HII regions, and they were used to establish a 
comparison baseline.  The remaining ten models were divided into five 
discrete flux levels ($f_{HII} =$ 0.05, 0.10, 0.15, 0.20, 0.25) with 
two models at each level in order to sample the same range of residual 
fluxes as seen in the real GSS galaxies.  This set of simulations 
contains 170 different galaxy models for a total of 2550 simulations. The 
galaxy models were created with the following structural 
parameters: m$_{F814W}(AB)=24.0$, $B/T=0.3$, $r_{e}=0\arcsecpoint 12$, 
$e=0.2$, $r_{d}=0\arcsecpoint 32$, $i=20\deg$, $\phi_b=\phi_d=60\deg$, 
and $n=4.0$.  These asymmetric image galaxy simulations were analyzed 
exactly the same way as the real data, and the biases in the parameter 
values recovered by GIM2D were then examined at each $f_{HII}$ 
level.  

Figures~\ref{err-i814-bt-asym-flux} and~\ref{err-rhl-rart-asym-flux} 
show the median systematic error on the recovered $B/T$, $r_{hl}$, and 
$I_{814}$ parameters as function of $f_{HII}$ in different six 
magnitude-size ranges.  Both the recovered total magnitude $I_{814}$ 
and half-light radius $r_{hl}$ show no trends with increasing 
$f_{HII}$, and the magnitude and size offsets in each magnitude-size 
range are in agreement with the systematic errors shown in 
Figures~\ref{err-map-i814-sepf} and~\ref{err-map-rhl-sepf}.  The bulge 
fraction is also fairly robust against asymmetries.  The median bulge 
fraction error is only about 0.1-0.2 for galaxies with $f_{HII} = 
0.20-0.25$.  However, increasing asymmetric flux leads to increasing 
scatter in the bulge fractions, and this scatter is skewed towards 
lower bulge fractions \ie, bulge fraction is always underestimated 
when asymmetries matter in a galaxy.  The asymmetry parameter 
$R_{A}+R_{T}$ (bottom half of Figure~\ref{err-rhl-rart-asym-flux}) 
recovers most of the asymmetric flux in big, bright galaxies.  Equally 
bright but smaller galaxies have measured $R_{A}+R_{T}$ slightly lower 
than big galaxies possibly due to the fact that the centroid of the 
bulge+disk models was allowed to vary by $\pm$ 1 pixel ($\pm 
0\arcsecpoint 15$) in the fits, and a shift in centroid would always 
be used by the fitting algorithm to reduce the overall amount of 
asymmetry ``seen'' by the smooth model.  The scatter in the recovered 
values of $R_{A}+R_{T}$ is higher for higher asymmetric fluxes due to 
model centroiding errors introduced by the asymmetries themselves.  
$R_{A}+R_{T}$ increasingly underestimates the asymmetric flux at 
fainter and fainter magnitudes as individual asymmetry sources become 
too faint to be picked out of the noise.

\section{Survey Selection Functions}\label{selecf}

Generalizing the formalism developed in \citet{simard99}, the 
observed distribution of galaxies in structural parameter space as a 
function of redshift $\Psi_{O}(\overline{{\bf W}},z)$ is the result of any 
inherent changes in the resident\footnote{It is very important to note 
the use of the term ``resident'' here and throughout the rest of the 
paper to refer to the intrinsic galaxy population at a given redshift 
$z$.  In the absence of real evolution in the galaxy population with 
redshift, all resident populations would be the same as the local 
population of galaxies.} galaxy distribution 
$\Psi_{U}(\overline{{\bf W}},z)$ in that space and of observational 
selection effects.  Selection effects are likely to be significant 
given the wide range of structural parameters observed locally 
\citep{bender92,burst97}.  It is therefore important to 
carefully characterize selection effects to disentangle them from real 
changes in $\Psi_{U}(\overline{{\bf W}},z)$.  The path from 
$\Psi_{U}(\overline{{\bf W}},z)$ to $\Psi_{O}(\overline{{\bf W}},z)$ is given by:

\begin{eqnarray}
\Psi_{O}(\overline{{\bf W}},z) = S_{PS} (\overline{{\bf W}},z) S_{UP} (\overline{{\bf W}},z) \Psi_{U} (\overline{{\bf W}},z), 
\label{seleq}
\end{eqnarray}

\noindent where $\overline{{\bf W}}$ is the full set of intrinsic structural 
parameters (note the use of lower and upper cases to distinguish 
between apparent and intrinsic structural parameter sets here).  The 
subscript $UP$ stands for ``Universe to Photometric sample,'' and the 
subscript $PS$ stands for ``Photometric sample to Spectroscopic 
sample.''  The resident galaxy distribution $\Psi_{U}(\overline{{\bf W}},z)$ 
is not known {\it a priori}.  Once the two selection functions in 
Equation~\ref{seleq} have been characterized, their product (denoted 
$S_{US}$ hereafter) shows the volume of the structural parameter space 
where real galaxies would have been observed if they existed in that 
region at high redshift.  The spectroscopic selection function $S_{PS} 
(\overline{{\bf W}},z)$ is derived in \citet{phillips02}, so the 
remainder of this section will focus on $S_{UP} (\overline{{\bf W}},z)$.

The selection function $S_{UP}(\overline{{\bf W}},z)$ contains the 
information needed to go from any sample of galaxies on the sky to the 
photometric catalog produced with SExtractor and reflects the adopted 
SExtractor detection parameters (detection threshold in sigmas, 
minimum detection area, etc.).  The detection thresholding method used 
by SExtractor depends critically on galaxy apparent surface 
brightness. The probability that a given object will be detected 
depends on total flux $F$, bulge fraction $B/T$, photobulge effective 
radius $r_e$, photobulge ellipticity $e$, photodisk scale length 
$r_d$, and photodisk inclination $i$.  For example, objects with 
larger $B/T$ will be easier to detect because they are more 
concentrated, and large objects will be harder to detect than smaller 
ones at a fixed total flux.  The selection function does not depend on 
the photobulge and photodisk position angles.  However, note that the 
selection function will also depend on disk internal extinction (if 
any), but this dependence is neglected here since GSS galaxies were 
analyzed with optically thin disks ($C_{abs} = 0$, recall 
Equation~\ref{othick-mag} in Section~\ref{surfmodel}).  In practice, 
$S_{UP}(\overline{{\bf W}},z)$ is derived from the selection function 
$S_{UP}(\overline{{\bf w}})$ determined as a function of the observed 
structural parameters.  The transformation $S_{UP}(\overline{{\bf w}}) 
\rightarrow S_{UP}(\overline{{\bf w}},z)$ can be made in each redshift bin 
using $k$-corrections calculated with the median observed galaxy 
$V_{606}-I_{814}$ color of the $B/T \leq 0.2$ galaxies at that 
redshift and the cosmological scale relations for the assumed 
cosmology.

$S_{UP}(\overline{{\bf w}})$ was constructed by generating 60,000 galaxy 
models with structural parameter values uniformly covering the ranges: 
$20.0$ $\leq$ $I_{814} \leq 25.0$, $0.0 \leq B/T \leq 1.0$, 
$0\arcsecpoint0 \leq r_{d} \leq 10\arcsecpoint 0$, $0 \leq$ sin $i 
\leq 0.9962$.  It is better here to uniformly generate disk inclinations in 
sin $i$ rather than in $i$ since randomly oriented, optically thin 
disks in space are expected to have a uniform sin $i$ distribution.  
Each model galaxy was added, one at a time, to an empty 20\arcsec 
$\times$ 20\arcsec section of a $F814W$ HST/GSS image (same image 
section as used in Section~\ref{sepfit}).  ``Empty'' here means that 
no objects were detected by SExtractor in that sky section with the 
same detection parameters used to construct the object catalog.  Using 
an empty section of the GSS ensured that $S_{UP}(\overline{{\bf w}})$ was 
constructed with the real background noise that was seen by the 
detection algorithm.  The background noise included read-out, sky and 
the brightness fluctuations of very faint galaxies below the detection 
threshold.  This last contribution to the background noise is 
particularly hard to model theoretically, and the current approach 
bypassed this problem.  SExtractor was run on each simulation with the 
same parameters that were used to build the SExtractor catalog.  The 
function $S_{UP}(\overline{{\bf w}})$ was taken to be the fraction of 
galaxies successfully detected and measured by SExtractor at each 
location $\overline{{\bf w}}$ in structural parameter space.

Figure~\ref{1d-sup} shows one-dimensional projections of 
$S_{UP}(\overline{{\bf w}})$ onto each of the six structural 
parameters ($F$, $B/T$, $r_e$, $e$, $r_d$, and $i$).  Different 
symbols show groups of galaxies with different bulge fractions.  
Clearly, some parameters are more important for the selection function 
than others, and bulge parameters are obviously more important for 
bulge-dominated galaxies and the same holds true for disk parameters.  
The selection function depends strongly on total apparent magnitude 
independent of bulge fraction, and bulge-dominated galaxies are more 
likely to be detected at all magnitudes than disk galaxies.  The 
selection function for bulge-dominated galaxies decreases with 
increasing bulge effective radius out to $r_{e} = 2\arcsec$ and 
remains relatively flat beyond that radius.  There is a very weak 
dependence of $S_{UP}(e)$ on bulge ellipticity.  The selection 
function for $B/T \ge 0.2$ galaxies is nearly independent of disk 
scale length whereas the selection function for $B/T < 0.2$ galaxies 
decreases out to $r_{d} = 2.2\arcsec$ and remains flat after that.  
There is virtually no dependence of $S_{UP}$ on the disk inclination 
angle, and this is somewhat surprising given that more inclined, optically thin 
disks should be easier to detect. This apparent puzzle was resolved by 
generating a set of 60,000 {\it pure disk} galaxy models and re-computing 
$S_{UP}$ with this new set. There was a clear dependence 
of $S_{UP}$ on disk inclination for this pure-disk galaxy 
set. The detectability of disks went from 0.55 at $i \sim 0\deg$ to 0.75 at 
$i \sim 80\deg$. Given that the disk sample in the selection function shown 
in Figure 15 includes galaxies with bulge fractions between 0.0 and 0.2, it 
appears that even a small (luminosity-wise) de Vaucouleurs bulge can boost 
the detectability of a galaxy enough to mask out the disk inclination 
dependence of the selection function.

\section{Comparison with The Medium Deep Survey}\label{mds}

The Medium Deep Survey (MDS, \citealp{ratnatunga99} and references 
therein) is the largest database of HST galaxy structural parameters 
in existence with 200,000 objects as of October 1998.  The images 
analyzed by the MDS team consist of MDS WFPC2 pure parallel 
observations as well as of HST archival observations of randomly 
selected WFPC2 fields such as the Groth Strip and the Hubble Deep Field 
among others.  The MDS team fitted the profiles of Groth Strip 
galaxies separately in $V_{606}$ and $I_{814}$ using a completely 
different analysis pipeline, a completely different likelihood maximization 
algorithm and a different bulge+disk model.  The parameters of the MDS 
bulge+disk galaxy model are sky background, $x$-$y$ centroid, 
orientation (bulge and disk are assumed to have the same position 
angle), bulge and disk axis ratios, bulge fraction $B/T$ and the ratio 
of the bulge/disk half-light radii.  The large sizes (thousands of objects 
each) of the MDS and DEEP/GIM2D Groth Strip structural catalogs make 
them ideally suited to run a check of one against the other. This is the first 
time that MDS results are compared against an independent work on 
such a scale.

The MDS Maximum-Likelihood Estimate (MLE) structural catalogs of the 
Groth Strip galaxies were extracted directly from the on-line MDS 
CD-ROMs, and the MDS catalogs were matched to the DEEP/GIM2D separate 
structural fit catalog using a matching radius of 0\arcsecpoint8 and 
a maximum $I_{814}$ magnitude difference of 1.  The match yielded 7138 
positive cross-identifications.  The results of the match are shown in 
Figure~\ref{mds-gim2d}.  The top left-hand panel shows the GIM2D 
galaxy model $I_{814}$ total magnitudes against the MDS galaxy model 
$I_{814}$ total magnitudes.  The long-dashed line is a one-to-one 
line, the filled circles are galaxies with 
$|I_{814,GIM2D}-I_{814,MDS}| \le 0.2$ mag, and the open circles are 
galaxies with magnitude differences larger than 0.2 mag.  The envelope 
of the data point distribution is clearly asymmetric with respect with 
the one-to-one line in its upper section \ie,  GIM2D magnitudes for 
some objects are too faint with respect to MDS magnitudes.  The 
asymmetry in the upper envelope is due to the fact that it is made up 
of two distributions.  One distribution comes from real photometric 
errors, and the distribution of its points as a function of distance from 
the one-to-one line is symmetric with respect to the lower envelope.  
The second distribution inside the upper envelope contains the most 
discrepant objects, and it is due to the fact that objects are more 
finely split in the DEEP/GIM2D structural catalog than in the MDS 
catalog.  Thus some single objects in the MDS catalog are two or more distinct 
objects in the DEEP/GIM2D catalog, and their DEEP/GIM2D 
magnitudes are therefore fainter.

The top right-hand panel of Figure~\ref{mds-gim2d} shows the GIM2D 
galaxy log half-light radii in arcseconds against the MDS radii for 
galaxies with $I_{814} \le 22$.  The long-dashed line is again the 
one-to-one line.  The half-light radii are in excellent agreement with 
no systematic differences.  The mean log radius difference 
(GIM2D$-$MDS) and rms scatter are (0.002, 0.049) for objects with 
$\Delta I_{814} \le 0.2$ mag (filled circles) and ($-$0.071, 0.168) for 
objects with $\Delta I_{814} > 0.2$ mag (open circles).  The rms scatter 
for the open circles is higher as one would expect from using different 
object splitting.

The GIM2D $I_{814}$ bulge fractions are compared against MDS $I_{814}$ 
bulge fractions for galaxies with $I_{814} \le 22$ in the last panel 
of Figure~\ref{mds-gim2d}.  The two vertical distributions at 
$(B/T)_{MDS} = 0.0$ (pure disk systems) and at $(B/T)_{MDS} = 1.0$ 
come from the MDS pipeline where objects below a certain S/N threshold 
and/or size are only fitted by a pure bulge or a pure disk model and 
not by the full bulge+disk model.  Setting these points aside for now, 
the mean $B/T$ difference (GIM2D$-$MDS) and rms scatter for 
intermediate $B/T$ galaxies are ($-$0.021,0.105) for the filled circles 
and ($-$0.046, 0.235) for the open circles.  The slight systematic 
offset and the increased rms scatter for the open circles is not 
surprising since objects that were split differently are likely to be 
classified differently.  The discrepancies between the MDS and GIM2D 
bulge fractions at $(B/T)_{MDS} = 0$ and $(B/T)_{MDS} = 1$ are best 
studied by looking at the two extremes where $(B/T)_{GIM2D} \ge 0.5$ 
at $(B/T)_{MDS} = 0$ and $(B/T)_{GIM2D} \le 0.5$ at $(B/T)_{MDS} = 1$.  
In the first case, the objects are unresolved in both MDS and GIM2D 
catalogs.  In the second case, there are different reasons behind the 
bulge fraction discrepancies.  Nineteen objects have $(B/T)_{GIM2D} 
\le 0.5$ at $(B/T)_{MDS} = 1$, and each was visually inspected.  Out 
of these 19 objects, 6 objects are irregular/peculiar galaxies or close pairs 
not separated in the MDS catalog, 2 are stars, 5 are single objects 
with $(B/T)_{MDS, I814} = 1$ and $(B/T)_{MDS, V606} \le 0.5$, 3 are 
equally well fit by the MDS and GIM2D models, 2 are single objects 
with knots in their MDS $I_{814}$ residual images, and 1 is a bad 
object match.  Even though these discrepant objects are very 
interesting, they represent a very small fraction of both catalogs, 
and the MDS and GIM2D results are generally in good agreement.

\section{Summary} \label{conc}

The structural parameters of galaxies in the Groth Survey Strip were 
measured from archival HST images as part of a combined HST and 
Keck/LRIS survey of the Strip by the DEEP team.  GSS galaxy surface 
brightness distributions were fitted with a 2D, PSF-convolved bulge+disk 
model using an implementation of the Metropolis algorithm to optimize 
model parameters.  A total of 7450 galaxies were fitted separately in 
$I_{814}$ and $V_{606}$.  648 galaxies with secure Keck/LRIS redshifts were 
also fitted simultaneously in both bandpasses.  The structural 
catalogs include image asymmetry parameters and rest-frame magnitudes 
and colors for bulge and disk components computed using two different 
sets of $k$-corrections.

This paper also provides full structural catalogs of two extensive sets of close to 6000 
simulations to allow detailed characterizations of the systematic and 
random errors in separate and simultaneous structural fits at any 
location in structural parameter space.  Error maps for galaxy 
magnitudes, half-light radii and bulge fractions are presented as 
examples.  The simultaneous structural fit simulations include varying 
bulge and disk colors and show that bulge and disk colors can 
be reliably measured down to $I_{814} (bulge) = 23.5$ and $I_{814} 
(disk) = 23.5$.  Similar formats are used for the real and the 
simulation catalogs so that interested users can study biases 
associated with different selection criteria to see which criteria 
will make the best use of the real data for the specific science goals 
being pursued.

The effects of unresolved, ``HII-region''-like asymmetric structures 
on fitting parameters were studied with a third set of 2550 
simulations.  Recovered total magnitudes, half-light radii and bulge 
fraction were, on average, robust against the presence of strong 
asymmetries with little or no systematic bias.  However, the scatter 
in the recovered parameters increased with increasing asymmetric flux, 
and this increased scatter was skewed towards lower bulge fractions.  
Bulge fractions are always underestimated when the presence of strong 
asymmetries matters.  The asymmetry parameter $R_{A}+R_{T}$ was found 
to be a good estimate of the total asymmetric flux present in large, 
bright galaxies.

The photometric selection function of the survey was mapped over a 
wide range of magnitudes, bulge fractions, and bulge/disk sizes to 
delineate the volume of structural space favored by the source 
detection algorithm.  The key quantity is surface brightness.  At a given 
magnitude, larger galaxies are harder to detect, and disks are harder to 
detect than bulges since disk profiles are less compact. There is little or 
no dependence of the selection function on bulge ellipticity and disk 
inclination.  The selection function, coupled with biases and errors 
from the simulations, gives a complete reproduction of the observational 
limits.

The structural parameters presented here were compared with the 
results from the Medium Deep Survey database. This is the first 
large scale comparison of MDS results against an independent source.  
The MDS and DEEP/GIM2D catalogs yields 7138 positive 
cross-identifications, and measurements of total magnitude, half-light 
radius and bulge fraction are all in good agreement with no 
systematic differences.  Bulge fraction classifications are in 
disagreement for a small fraction of the galaxies.

The combination of the three essential ingredients of quantitative 
galaxy morphology (very large sets of structural measurements, 
detailed characterization of biases and errors, and mapping of the 
multidimensional photometric selection function) found in this paper 
is ideal to pursue direct comparisons between observations and the 
latest models of galaxy formation and evolution.

\acknowledgments

The development of GIM2D greatly benefited from the experiences of a 
core user group which includes Kim-Vy Tran, Brad Holden, Francine 
Marleau, Katherine Wu, Sasha Hinkley, and Daniel McIntosh.  L.S. is 
indebted to them for their help.  Special thanks to Katherine Wu and 
Roelof de Jong for their very thorough readings of the manusript. L. S. 
acknowledges many insightful discussions with P. D. Simard that 
significantly improved this paper.  L. S. also gratefully acknowledges 
partial financial support from the Natural Sciences and Engineering 
Research Council of Canada through the award of a Postdoctoral 
Fellowship.  This work was  supported by NSF grants AST95-29098 
and 00-71198 and NASA/HST grants AR-07532.01,  AR-06402.01 
and AR-05801.01.

\clearpage
\figcaption[figure1.ps]{$HST/WFPC2/F814W$ image of the Groth Survey 
Strip Field 8/Wide Field Camera Chip 3. Image scale is 0\arcsecpoint 1 pixel$^{-1}$, 
and the total exposure time is 4400 seconds. \label{gss-science-image}}

\figcaption[figure2.ps]{SExtractor segmentation image of GSS image in 
Figure~\ref{gss-science-image}. Every pixel in this image was assigned 
the non-zero SExtractor flag value of its parent object or a flag 
value of zero if it was a background pixel. GIM2D uses this segmentation 
image to independently fit objects which are very close to each other.
\label{gss-segmen-image}}

\figcaption[figure3.ps]{Mosaics of GSS thumbnail images drawn {\it at 
random} from the DEEP/GSS sample.  The extracted region for each 
galaxy is twenty times the area of its 1.5$-\sigma_{bkg}$ isophote.  {\it 
First mosaic - Science thumbnail images}:  The numbers shown around 
each science thumbnail image are the DEEP/GSS object ID (top left), 
DEEP Keck redshift (top right), observed galaxy total magnitude 
$I_{814}$ (bottom left) and observed galaxy $V_{606}-I_{814}$ color 
(bottom right). {\it Second mosaic - Output PSF-convolved GIM2D model 
thumbnail images}: The numbers shown around each model thumbnail image 
are the DEEP/GSS object ID (top left), bulge fraction $B/T$ (top 
right), rest-frame $B$-band galaxy total absolute magnitude (bottom 
left), and galaxy log half-light radius in arcseconds (bottom right).  
{\it Third mosaic - Residual thumbnail images (science $-$ model)}: 
Numbers shown around each residual thumbnail image are the DEEP/GSS ID 
(top left), $RT3+RA3$ asymmetry parameter (top right), concentration 
index $C$ (bottom left) and asymmetry index $A$ (bottom right).  The 
same greyscale was used for the science, model and residual thumbnail 
images of a given galaxy, but a different greyscale was used for each 
galaxy.  The bottom greyscale value for each galaxy was set to be 
5-$\sigma_{bkg}$ below the science thumbnail background level, and the 
top greyscale value was set to be 10-$\sigma_{bkg}$ above the background.  
Notice the close galaxy pair 192\_{2330} which was successfully 
deblended by SEXtractor and independently fitted by GIM2D. Also notice the 
wealth of interesting structures in all the residual images.
\label{gss-image-mosaics}}

\figcaption[figure4.ps]{Two-dimensional maps of systematic and random 
galaxy magnitude errors from 5995 DEEP/GSS separate structural fit 
simulations.  {\it Top left-hand panel}: Systematic error on 
recovered galaxy total magnitude $I_{814,rec}$ as a function of {\it input} 
galaxy log half-light radius $r_{hl,input}$ in arcseconds and 
{\it input} galaxy total magnitude $I_{814,input}$.  The top number in 
each cell is the mean magnitude error ($I_{814,rec} - I_{814,input}$), 
and the bottom number is the number of simulations created 
in that cell.  {\it Top right-hand panel}: 1-$\sigma$ random error on 
$I_{814,rec}$ ($\sigma$($I_{814,rec}-I_{814,input}$)) as a function of 
log $r_{hl,input}$ and $I_{814,input}$.  {\it Bottom left-hand panel}: 
Systematic error on $I_{814,rec}$ as a function of {\it recovered} 
galaxy log half-light radius $r_{hl,rec}$ in arcseconds and 
$I_{814,rec}$.  {\it Bottom right-hand panel}: Actual DEEP/GSS 
$I_{814,obs}$-log $r_{hl,obs}$ observations.  Since recovered 
quantities are the simulation equivalent of the observed ones, this 
panel should be directly compared to the bottom left-hand panel to see 
how important magnitude errors are in different regions of the 
observational magnitude-size plane.  Although such 2D maps can be very 
useful tools, one should keep in mind that they are only two-dimensional 
projections of a complex multi-dimensional error function.
\label{err-map-i814-sepf}}

\figcaption[figure5.ps]{Same as Figure~\ref{err-map-i814-sepf} except 
that log half-light radius errors are shown here.
\label{err-map-rhl-sepf}}

\figcaption[figure6.ps]{Same as Figure~\ref{err-map-i814-sepf} except 
that bulge fraction errors are shown here.
\label{err-map-bt-sepf}}

\figcaption[figure7.ps]{Two-dimensional maps of systematic and random 
galaxy bulge fraction errors from DEEP/GSS separate structural fit 
simulations with input galaxy half-light radius $r_{hl,input} \ge 
0 \arcsecpoint 1$.  {\it Top left-hand panel}: Systematic error on 
recovered galaxy bulge fraction $(B/T)_{rec}$ as a function of {\it input} 
galaxy bulge fraction $(B/T)_{input}$ and {\it input} galaxy total magnitude 
$I_{814,input}$.  The top number in each cell is the mean bulge fraction error ($(B/T)_{rec} - (B/T)_{input}$), 
and the bottom number is the number of simulations created 
in that cell.  {\it Top right-hand panel}: 1-$\sigma$ random error on 
$(B/T)_{rec}$ ($\sigma$($(B/T)_{rec}-(B/T)_{input}$)) as a function of 
$(B/T)_{input}$ and $I_{814,input}$.  {\it Bottom left-hand panel}: 
Systematic error on $(B/T)_{rec}$ as a function of $(B/T)_{rec}$ and 
$I_{814,rec}$.  {\it Bottom right-hand panel}: Actual DEEP/GSS 
$(B/T)_{obs}$-$I_{814,obs}$ observations.  Since recovered 
quantities are the simulation equivalent of the observed ones, this 
panel should be directly compared to the bottom left-hand panel to see 
how important bulge fraction errors are in different regions of the 
observational magnitude-bulge fraction plane.
\label{btm-btbias-sepf}}

\figcaption[figure8.ps]{Two-dimensional maps of systematic and random 
galaxy magnitude errors from 5195 DEEP/GSS simultaneous structural fit 
simulations.  {\it Top left-hand panel}: Systematic error on 
recovered galaxy total magnitude $I_{814,rec}$ as a function of 
{\it input} galaxy log half-light radius $r_{hl,input}$ in arcseconds and 
{\it input} galaxy total magnitude $I_{814,input}$.  The top number 
in each cell is the mean magnitude error ($I_{814,rec} - I_{814,input}$), 
and the bottom number is the number of simulations created 
in that cell.  {\it Top right-hand panel}: 1-$\sigma$ random error on 
$I_{814,rec}$ ($\sigma$($I_{814,rec}-I_{814,input}$)) as a function of 
log $r_{hl,input}$ and $I_{814,input}$.  {\it Bottom left-hand panel}: 
Systematic error on $I_{814,rec}$ as a function of {\it recovered} 
galaxy log half-light radius $r_{hl,rec}$ in arcseconds and 
$I_{814,rec}$.  {\it Bottom right-hand panel}: Actual DEEP/GSS 
$I_{814,obs}$-log $r_{hl,obs}$ observations.  The number of data 
points is smaller than in 
Figures~\ref{err-map-i814-sepf},~\ref{err-map-rhl-sepf}, 
and~\ref{err-map-bt-sepf} since simultaneous bandpass fits were 
performed only on GSS galaxies with secure DEEP Keck redshifts.  Since 
recovered quantities are the simulation equivalent of the observed 
ones, this panel should be directly compared to the bottom left-hand 
panel to see how important magnitude errors are in different regions of 
the observational magnitude-size plane.  Although such 2D maps can be 
very useful tools, one should keep in mind that they are only 
two-dimensional projections of a complex multi-dimensional error 
function.
\label{err-map-i814-simulf}}

\figcaption[figure9.ps]{Same as Figure~\ref{err-map-i814-simulf} except 
that log half-light radius errors are shown here.
\label{err-map-rhl-simulf}}

\figcaption[figure10.ps]{Same as Figure~\ref{err-map-i814-simulf} except 
that bulge fraction errors are shown here.
\label{err-map-bt-simulf}}

\figcaption[figure11.ps]{Structural component colors recovered by 
GIM2D from simultaneous structural fit simulations. 
{\it Top three panels}: Recovered galaxy $V_{606}-I_{814}$ color 
versus input galaxy $V_{606}-I_{814}$ color for input galaxy total 
magnitudes in the ranges $20.0 \le I_{814} < 21.0$, $21.0 \le 
I_{814} < 22.0$, and $22.0 \le I_{814} < 23.5$. The mean and rms color differences 
(recovered$-$input) are (0.002, 0.020), (0.010, 0.034), and (0.063, 
0.087) for the three ranges respectively. $V_{606, lim}$ = 26.0. 
{\it Middle three panels}: Recovered bulge $V_{606}-I_{814}$ color 
versus input bulge $V_{606}-I_{814}$ color for input bulge magnitudes in the ranges 
$20.0 \le I_{814}{\rm (bulge)} < 21.0$, $21.0 \le I_{814} {\rm (bulge)} < 22.0$, and 
$22.0 \le I_{814} {\rm (bulge)} < 23.5$. The mean and rms bulge color differences 
(recovered$-$input) are (-0.001, 0.255), (0.022, 0.120), and (0.034, 
0.186) for the three ranges respectively. $V_{606, lim} {\rm (bulge)}$ = 
26.0, $r_{hl,814} \ge 0\arcsecpoint 15$, and $r_{hl,606} \ge 0\arcsecpoint 15$.
{\it Bottom three panels}: Recovered disk $V_{606}-I_{814}$ color 
versus input disk $V_{606}-I_{814}$ color for input disk magnitudes in the ranges 
$20.0 \le I_{814}{\rm (disk)} < 21.0$, $21.0 \le I_{814} {\rm (disk)} < 22.0$, and 
$22.0 \le I_{814} {\rm (disk)} < 23.5$. The mean and rms disk color differences 
(recovered$-$input) are (0.008, 0.047), (0.006, 0.211), and (0.030, 
 0.304) for the three ranges respectively. $V_{606, lim} {\rm (disk)}$ = 
26.0, $r_{hl,814} \ge 0\arcsecpoint 15$, and $r_{hl,606} \ge 0\arcsecpoint 15$.
\label{rec-inp-bulge-disk-colors}}

\figcaption[figure12.ps]{Two-dimensional maps of systematic galaxy bulge 
fraction errors from DEEP/GSS separate and simultaneous structural fit 
simulations as a function of {\it input} galaxy bulge fraction and {\it input} 
bulge effective radius/disk scale length ratio.  {\it Left-hand panel}: Systematic error on 
recovered galaxy bulge fraction $(B/T)_{rec}$ from separate 
structural fits (SPF) as a function of {\it input} 
galaxy bulge fraction $(B/T)_{input}$ and {\it input} log bulge 
effective radius/disk scale length ratio $r_{e}/r_{d}$.  The top number in each cell is the 
mean bulge fraction error ($(B/T)_{rec} - (B/T)_{input}$), the middle 
number in each cell is the minimum error, and the bottom number in each 
cell is the maximum error.  {\it Right-hand panel}: Same as left-hand 
panel but for simultaneous structural fits. These maps show that there 
do not seem to be any regions where GIM2D can systematically mistake 
disks for bulges and vice versa.
\label{bt-irr-nobias}}

\figcaption[figure13.ps]{Median systematic errors on recovered bulge fraction $B/T$ 
and galaxy total apparent magnitude $I_{814}$ as a function of asymmetric flux 
fraction $f_{HII}$.  {\it Top six panels}: Median $B/T$ systematic error in six different 
magnitude-size ranges. {\it Bottom six panels}: Median $I_{814}$ systematic error 
in six different magnitude-size ranges. Vertical error bars are 68$\% $ lower and 
upper bounds.
\label{err-i814-bt-asym-flux}}

\figcaption[figure14.ps]{{\it Top six panels}:  Median systematic 
error on recovered galaxy log half-light radius $r_{hl}$ as a function of asymmetric flux fraction $f_{HII}$ in 
six different magnitude-size ranges.  Vertical error bars are 68$\% $ 
lower and upper bounds.  {\it Bottom six panels}:  Median recovered 
$R_{A}+R_{T}$ asymmetry index values as a function of asymmetric flux 
fraction $f_{HII}$ in six different magnitude-size ranges.  
Vertical error bars are 68$\% $ lower and upper bounds.
\label{err-rhl-rart-asym-flux}}

\figcaption[figure15.ps]{Six one-dimensional projections of the 
photometric selection function $S_{UP}(\overline{{\bf w}})$: galaxy 
total apparent magnitude $I_{814}$ ({\it top left}), galaxy bulge 
fraction $(B/T)$ ({\it top middle}), bulge effective radius $r_{e}$ 
in arcseconds ({\it top right}), bulge ellipticity $e$ ({\it bottom 
left}), disk scale length $r_{d}$ in arcseconds ({\it bottom 
middle}), and disk inclination angle $i$ ({\it bottom right}). Filled 
circles are simulated galaxies with $0.0 \le (B/T) \le 0.2$, pluses 
are galaxies with $0.2 < (B/T) \le 0.8$, and triangles are galaxies 
with $0.8 < (B/T) \le 1.0$.
\label{1d-sup}}

\figcaption[figure16.ps]{Comparison between MDS and GIM2D parameters 
for Groth Strip galaxies.  {\it Top left}: GIM2D versus MDS total 
galaxy model $I_{814}$ magnitudes. {\it Top right}: GIM2D versus MDS galaxy 
semi-major axis log half-light radiii in arcseconds for galaxies with 
$I_{814} \le 22$.  {\it Lower left}: GIM2D versus MDS galaxy bulge 
fractions for galaxies with $I_{814} \le 22$.  Filled circles are galaxies with 
$|I_{814,GIM2D}-I_{814,MDS}| \le 0.2$, and open circles are the 
remainder of the sample. The long dashed lines in all three panels are 
one-to-one lines.\label{mds-gim2d}}

\clearpage 
\begin{deluxetable}{ccccrcl}
\tablewidth{0pt}
\tablecaption{HST/WFPC2/GSS Image Datasets}
\tabletypesize{\footnotesize}
\tablehead{
\colhead{DEEP/GSS} & \colhead{HST Target} & \colhead{$\alpha$} & 
\colhead{$\delta$} & \colhead{Observation} & \colhead{PI} & 
\colhead{MDS}\\
\colhead{Field ID} & \colhead{Name} & \colhead{(J2000.0)} & 
\colhead{(J2000.0)} & \colhead{Date} & \colhead{} & \colhead{Field ID}
}
\startdata
4 & FIELD-141803+523 & 14:18:03 & +52:32:10 & 16 March 1994 & Groth & u26x4\\
5 & FIELD-141757+523 & 14:17:56 & +52:31:00 & 11 March 1994 & Groth & u26x5\\
6 & FIELD-141750+522 & 14:17:50 & +52:29:51 & 13 March 1994 & Groth & u26x6\\
7 & DEEP-SURVEY FIELD-2 & 14:17:43 & +52:28:41 & 22 March 1994 & Westphal & u2ay1\\
8 & FIELD-141737+522 & 14:17:37 & +52:27:31 & 10 March 1994 & Groth & u26x7\\
9 & FIELD-141731+522 & 14:17:30 & +52:26:21 & 11 March 1994 & Groth & u26x8\\
10 & FIELD-141724+522 & 14:17:24 & +52:25:11 & 16 March 1994 & Groth & u26x9\\
11 & FIELD-141717+522 & 14:17:17 & +52:24:01 & 07 March 1994 & Groth & u26xa\\
12 & FIELD-141711+522 & 14:17:10 & +52:22:51 & 07 March 1994 & Groth & u26xb\\
13 & FIELD-141704+522 & 14:17:04 & +52:21:41 & 10 March 1994 & Groth & u26xc\\
14 & FIELD-141658+522 & 14:16:57 & +52:20:31 & 13 March 1994 & Groth & u26xd\\
15 & FIELD-141651+521 & 14:16:51 & +52:19:22 & 16 March 1994 & Groth & u26xe\\
16 & FIELD-141645+521 & 14:16:44 & +52:18:11 & 17 March 1994 & Groth & u26xf\\
17 & FIELD-141638+521 & 14:16:38 & +52:17:01 & 14 March 1994 & Groth & u26xg\\
18 & FIELD-141632+521 & 14:16:31 & +52:15:51 & 16 March 1994 & Groth & u26xh\\
19 & FIELD-141626+521 & 14:16:25 & +52:14:41 & 17 March 1994 & Groth & u26xi\\
20 & FIELD-141619+521 & 14:16:19 & +52:13:31 & 10 March 1994 & Groth & u26xj\\
21 & FIELD-141613+521 & 14:16:12 & +52:12:21 & 09 March 1994 & Groth & u26xk\\
22 & FIELD-141606+521 & 14:16:06 & +52:11:11 & 17 March 1994 & Groth & u26xl\\
23 & FIELD-141600+521 & 14:15:59 & +52:10:01 & 09 March 1994 & Groth & u26xm\\
24 & FIELD-141553+520 & 14:15:53 & +52:08:51 & 15 March 1994 & Groth & u26xn\\
25 & FIELD-141547+520 & 14:15:46 & +52:07:40 & 
09\thinspace\thinspace\thinspace\thinspace\thinspace April 1994 & Groth & u26xo\\
26 & FIELD-141540+520 & 14:15:40 & +52:06:30 & 08 March 1994 & Groth & u26xp\\
27 & FIELD-141534+520 & 14:15:33 & +52:05:20 & 08 March 1994 & Groth & u26xq\\
28 & FIELD-141527+520 & 14:15:27 & +52:04:10 & 08 March 1994 & Groth & u26xr\\
29 & FIELD-141823+523 & 14:15:21 & +52:02:59 & 15 March 1994 & Groth & u26x1\\
30 & FIELD-141816+523 & 14:15:14 & +52:01:49 & 15 March 1994 & Groth & u26x2\\
31 & FIELD-141810+523 & 14:15:08 & +52:00:39 & 02\thinspace\thinspace\thinspace\thinspace\thinspace April 1994 & Groth & u26x3\\
\enddata
\label{hst-images-lst}
\end{deluxetable}

\end{document}